\documentclass[
reprint,
superscriptaddress,
amsmath, 
amssymb
]{revtex4-2}
\usepackage{graphicx}
\usepackage{dcolumn}
\usepackage{bm}
\usepackage{hyperref}
\begin{document}

\preprint{APS/123-QED}

\title{Inverted loss engineering in functional material covered waveguides}

\author{Ayvaz I. Davletkhanov}
\affiliation{Skolkovo Institute of Science and Technology, Moscow 143026, Russia}
\author{Aram A. Mkrtchyan}
\affiliation{Skolkovo Institute of Science and Technology, Moscow 143026, Russia}
\author{Dmitry A. Chermoshentsev}
\affiliation{Skolkovo Institute of Science and Technology, Moscow 143026, Russia}
\author{Mikhail V. Shashkov}
\affiliation{Boreskov Institute of Catalysis SB RAS, Novosibirsk 630090, Russia}
\author{Daniil A. Ilatovskii}
\affiliation{Skolkovo Institute of Science and Technology, Moscow 143026, Russia}
\author{Dmitry V. Krasnikov}
\affiliation{Skolkovo Institute of Science and Technology, Moscow 143026, Russia}
\author{Albert G. Nasibulin}
\affiliation{Skolkovo Institute of Science and Technology, Moscow 143026, Russia}
\author{Yuriy G. Gladush}
\email{Y.Gladush@skoltech.ru}
\affiliation{Skolkovo Institute of Science and Technology, Moscow 143026, Russia}

\begin{abstract}

Optical waveguides, covered with thin films, which transmittance can be controlled by external action, are widely used in various applications from optical modulators to saturable absorbers. It is natural to suggest that the waveguide losses will be proportional to the covering material absorption. We demonstrate that under certain conditions this simple assumption fails. Instead, we observe the reduction of the film material absorption can lead to an increase in the waveguide propagation losses. For this, we use a side polished fiber covered with a single-walled carbon nanotube thin film whose absorption is attenuated either due to saturable absorption or electrochemical gating. For the films thicker than 50 nm, we observe saturable absorption to turn into light induced absorption with nonmonotonic dependence on the incident power. With a numerical simulation and analytical approach, we identify that this nontrivial behavior comes from mode reshaping and predict required parameters for its observation.  

\end{abstract}

\maketitle

\section*{Introduction}

Covering waveguides with functional materials can bring new functionality to the otherwise passive optical elements. Examples include graphene and related 2D materials, carbon nanotubes, vanadium dioxide, topologic insulators, and many others used in integrated photonics for modulators, optical switches, photodetectors, and nanoheaters \cite{Romagnoli2018,Youngblood2014,Bonaccorso2010,Bao2012,Jeong2013,Tan2020,Sun2016,Yu2017,Parra2021,Janjan2017}. Another important platform utilizes optical fibers when functional materials cover either a side-polished fiber (SPF) or a tapered fiber to produce sensors \cite{Khan2014,Hu2018,An2020,Cao2019,Yan2015,Zheng2015,Abouraddy2007}, modulators \cite{Wang2015,Chen2015,Li2014,Liu2013,Heidari2021,AlexanderSchmidt2016,Yu2013}, polarizers \cite{Heidari2021,Kou2014,Bao2011,Li2021,Chu2017,Nikbakht2019}, or saturable absorbers \cite{Li2015, Song2007,Jeong2016,Lee2017,Lee2015,Zhao2012}. The efficiency of these devices directly depends on the light and material interaction strength, which is defined by the overlap integral of the covering material and the evanescent tail of the waveguide mode. Several approaches have been proposed to increase the overlap integral by adjusting the waveguide geometry and the material coating technique \cite{Zapata2016,Jiang2022,Chen2021,Demongodin2019,Sederberg2014}. For example, an overlayer with a higher refractive index pulls the mode towards itself, which can significantly increase the overlap integral \cite{Li2021,Chu2017,Zapata2016,Park2015}. The variation of the optical parameters of the covering material affect the shape of the mode that enables to control the overlap integral, i.e., the interaction strength. Many of the mentioned above applications, like optical switches and modulators, optical limiters, and saturable absorbers, rely on the change of the imaginary part of the refractive index, while the variation of its real part remains small. The imaginary part variation can also change the overlap integral and consequently the losses through the waveguide.

In this work, we demonstrate theoretically and experimentally that a continuous reduction of the imaginary part of the refractive index of the covering material not necessarily leads to a proportional decrease in the losses through the waveguide but can even cause an increase in the waveguide absorbance. With a help of numerical simulations, we show that this effect can be attributed to the mode reshaping and change in the overlap integral value. For experimental demonstration, we use a practically important system of a polymer-free single-walled carbon nanotube (SWCNT) thin film covering the SPF, which is used for the mode-locking of fiber lasers \cite{Mkrtchyan2019}. We apply two experimental methods to decrease the material absorption: an electrochemical gating and resonant absorption saturation. In both cases, we find that it leads to the absorbance increase through the waveguide for sufficiently thick film in accordance with numerical simulations. Finally, we show that this effect can be resembled by a simplified analytical approach that allows us to predict the conditions, under which it can be observed. 

\section*{Results}

\begin{figure*}
\includegraphics[width=\linewidth,keepaspectratio]{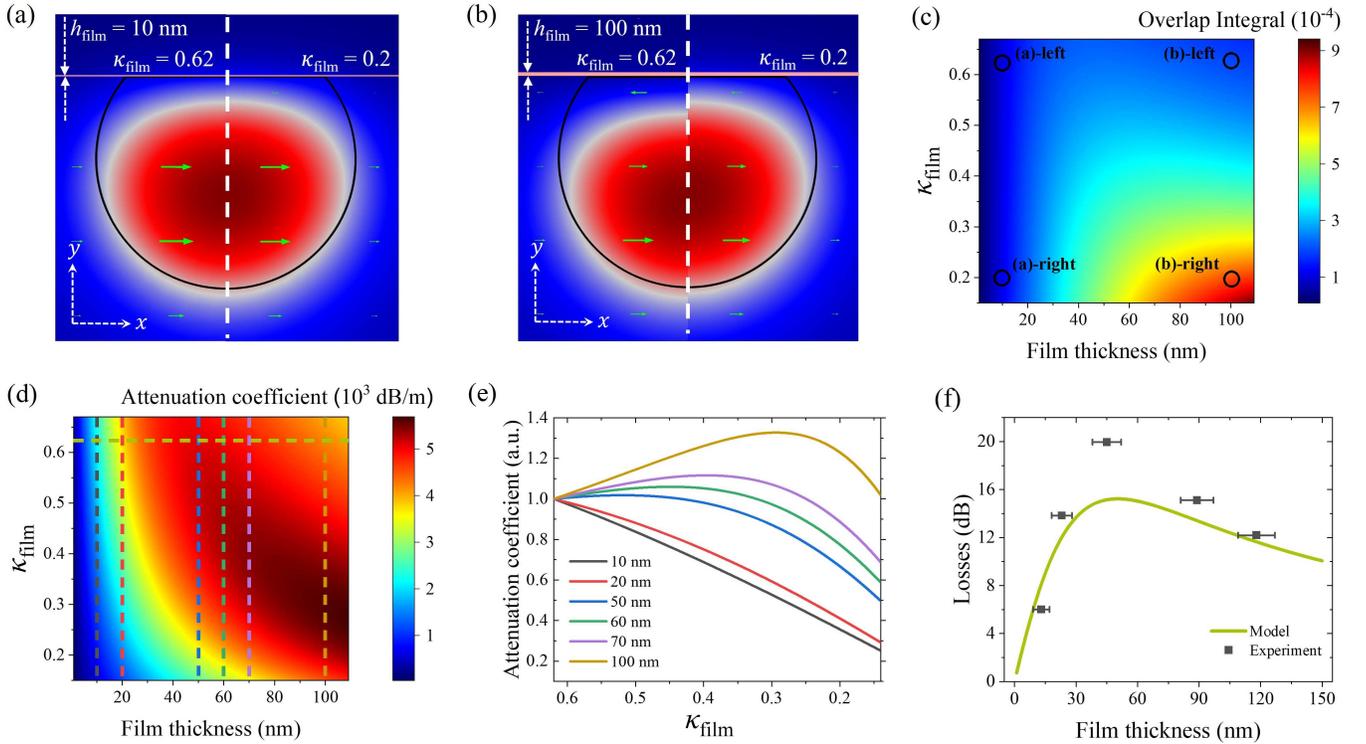}
\caption{\label{fig:fig1}(a, b) Cross-sectional electric field amplitude distributions of the TE modes in the SPF for selected SWCNT film thicknesses $h_{\textrm {film}}$ = 10 and 100 nm and SWCNT film imaginary part of refractive indexes $\kappa_{\textrm {film}}$ = 0.62 and 0.2. (c) Contour plot of the overlap integral of the TE mode and the SWCNT film on the SPF. (d) Contour plot of the attenuation coefficient of the SPF with covered SWCNT film. Vertical cut-lines are shown in plot (e), horizontal – in plot (f). (e) Dependence of the attenuation coefficient on $h_{\textrm {film}}$ normalized to the initial value of $\kappa_{\textrm {film}}$=0.62 for selected film thicknesses. (f) Modeled dependence of the attenuation on unsaturated SWCNT film thickness (line) and its experimentally measured values (dots).}
\end{figure*}

We investigate the SPF with a thin film of SWCNTs covered by a liquid overlayer. The details of the sample preparation can be found in \hyperref[methods]{Methods} section. In brief, for SWCNTs, we implement an aerosol floating catalyst synthesis method that allows to obtain a thin film of pure uniform randomly oriented carbon nanotube network with a required thickness without any liquid chemistry steps \cite{Mustonen2015,Tian2011}. The mean diameter of the SWCNTs is tuned during the synthesis to 1.4 nm \cite{Nasibulin2006}, corresponding to $S_{11}$ interband transition at 1.55 $\mu$m, which matched with the wavelength of the erbium-doped fiber laser (see the absorbance spectrum of carbon nanotubes in \hyperref[suppinfo]{Figure~S1 in Supplementary Information}). We synthesized a series of SWCNT films with various thicknesses in the range from 10 to 120 nm, which were controlled using AFM (see \hyperref[suppinfo]{Figure~S2}), similar to our previous work \cite{Ermolaev2020}. Then, SWCNT films are transferred onto the SPF by a dry-transfer technique \cite{Mkrtchyan2019,Gladush2019} and covered by index-matched liquid to enhance their interaction with light \cite{Li2021,Chu2017,Zapata2016,Park2015}. As an index-matched overlayer, we use ionic liquid Bmim $\textrm{NTf}_2$ with refractive index $n_{\textrm {IL}}$=1.41 and no absorption at 1.55 $\mu$m wavelength. In addition, we use it to modify the absorption of the carbon nanotube film by electrochemical gating \cite{Kataura1999}. To fabricate the ionic cell we use the same design as in previous work \cite{Gladush2019}. With no applied voltage ionic liquid exclusively plays the role of the index-matched overlayer.

\subsection*{Light propagation simulation}

First, we numerically investigate the light propagation through the SPF covered with the SWCNT film for various film thicknesses. Interband transitions require polarization along carbon nanotubes \cite{Ichida2004}, which are oriented predominantly in the plane of the film. So, we are interested mostly in the polarization in the plane with a polished surface, which we further refer to as TE. The parameters used for simulation are summarized in \autoref{tab:table1}. Measurements of refractive indices of the SWCNT film and the ionic liquid overlayer are described in \hyperref[methods]{Methods}.

The absorption coefficient of the SPF covered with SWCNTs depends on the imaginary part of the refractive index of the film, $\kappa_\textrm{film}$, related to its material losses as $\alpha_\textrm{film}=\frac{4\pi}{\lambda}\kappa_\textrm{film}$ , and the overlap integral of the mode profile of the carbon nanotube film:

\begin{eqnarray} 
\alpha_\textrm{SPF}=\frac{4\pi}{\lambda}
\frac{\iint_{-\infty}^{+\infty}\kappa(x,y){|E(x,y)|}^2 \mathrm{d}x \mathrm{d}y}
{\iint_{-\infty}^{+\infty}{|E(x,y)|}^2 \mathrm{d}x \mathrm{d}y}=\nonumber\\
 =\frac{4\pi}{\lambda}\kappa_\textrm{film} 
\frac{\iint_\textrm{over film}{|E(x,y)|}^2 \mathrm{d}x \mathrm{d}y}
{\iint_{-\infty}^{+\infty}{|E(x,y)|}^2 \mathrm{d}x \mathrm{d}y}=\nonumber\\
 =\frac{4\pi}{\lambda}\kappa_\textrm{film} \times \textrm{Overlap Integral}.
\label{eq:1}
\end{eqnarray}

\begin{table}[b]
  \caption{Parameters for simulation the light propagation through the SPF covered with the SWCNT film}
  \label{tab:table1}
  \begin{tabular}{ll}
    \hline
    Parameter & Value  \\
    \hline
    Wavelength [$\mu$m]  & 1.55  \\
    Core radius [$\mu$m]  & 4.5  \\
    Refractive index of the core  & 1.450  \\
    Refractive index of the cladding  & 1.445  \\
    Refractive index of overlayer $n_\textrm{IL}$  & 1.41  \\
    Real part of SWCNT refractive index $n_\textrm{film}$  & 1.43  \\
    Imaginary part of SWCNT refractive index $\kappa_\textrm{film}$ & 0.62  \\
    \hline
  \end{tabular}
\end{table}

\noindent Here, $\lambda$ is the wavelength of light and we neglect the losses of the fiber and ionic liquid. We consider $\kappa_{\textrm {film}}$=0.62 for the pristine nanotube film and model how the field distribution in SPF-SWCNT changes when $\kappa_{\textrm {film}}$ value decreases. The results for relatively thin (10 nm) and thick (100 nm) films are shown in \autoref{fig:fig1}a and \autoref{fig:fig1}b, respectively. First of all, we see that the presence of the SWCNT film deforms the mode towards the opposite side from the polished surface. For the thin film, the field profile does not change as $\kappa_{\textrm {film}}$ reduces from 0.62 to 0.2. In contrast, for the thick film, the field distribution is shifted further from the surface for larger values of $\kappa_{\textrm {film}}$, and moves closer, increasing the overlap integral, if $\kappa_{\textrm {film}}$ is decreased. The difference can be explicitly seen in the electric field amplitude distributions of the mode along the central cut lines (\hyperref[suppinfo]{Figure~S5 in Supplementary Information}). It leads to two competing processes: the reduction of $\kappa_{\textrm {film}}$ decrease the absorption, while the increase in the overlap integral increases it. To understand how it affects the losses of the SPF-SWCNT, we plot the mode-film overlap integral \autoref{fig:fig1}c, points with parameters corresponding to \autoref{fig:fig1}(a, b) are highlighted) and attenuation coefficient $\alpha_\textrm{SPF}$ (see \autoref{fig:fig1}d) following \autoref{eq:1} as a function of film thickness and $\kappa_{\textrm {film}}$. In \autoref{fig:fig1}c we can clearly notice that for thin films overlap integral does not depend on $\kappa_{\textrm {film}}$, while for thicker films it grows significantly for smaller $\kappa_{\textrm {film}}$, which is the consequence of the mode reshaping. The resulting attenuation coefficient in \autoref{fig:fig1}d shows monotonic behavior for smaller thicknesses. However, for thicknesses more than 50 nm we can define two regions — where the attenuation coefficient decreases with a decrease in $\kappa_{\textrm {film}}$ $(\frac{\mathrm{d}\alpha_\textrm{SPF}}{\mathrm{d}\kappa_\textrm{film}} > 0)$ and opposite behavior $(\frac{\mathrm{d}\alpha_\textrm{SPF}}{\mathrm{d}\kappa_\textrm{film}} < 0)$ separated by a point where $\alpha_\textrm{SPF}$ reaches the maximum for a given film thickness $(\frac{\mathrm{d}\alpha_\textrm{SPF}}{\mathrm{d}\kappa_\textrm{film}} = 0)$.

For a better visibility in \autoref{fig:fig1}e, we demonstrate the behavior of the attenuation coefficients, normalized to initial values of $\kappa_{\textrm {film}}$, for various film thicknesses corresponding to vertical cut-lines in \autoref{fig:fig1}d. Here, we observe that for 10 and 20 nm films, the attenuation coefficient decreases almost linearly with the reduction of $\kappa_{\textrm {film}}$, as expected, since the overlap integral does not change and does not contribute to the losses of the waveguide. Starting from 50 nm thick films, the slope of the attenuation coefficient curve changes the sign to positive, leading to an increase of the losses with the decrease of the $\kappa_{\textrm {film}}$. Under a certain value of $\kappa_{\textrm {film}}$ the absorption coefficient reaches the maximum and then starts to decrease because the growing overlap integral cannot compensate for the reduction of material losses when the $\kappa_{\textrm {film}}$ gets sufficiently small. We note that this type of nonmonotonic attenuation coefficient behavior cannot be observed by variation of the real part of the refractive index of SWCNTs in the physically reachable range, see \hyperref[suppinfo]{Figure~S6 in Supplementary Information}.

\subsection*{Experimental demonstration}

As the first experimental verification, we measure small-signal losses of SWCNT films on the SPF with identical $\kappa_{\textrm {film}}$=0.62, but different thicknesses, using the low-power CW laser (10 mW) to eliminate the contribution of nonlinear losses. According to the simulation, the SPF-SWCNT losses should have a maximum at the film thickness of 50 nm, see green line in \autoref{fig:fig1}f. The experimental results are shown as a black point on the plot, where the film thickness was measured by the AFM. We see that in the experiment the losses follow the same trend having the maximum for the film thickness of around 50 nm. The higher losses in the experiment we attribute to additional losses when the mode evolves from symmetric on the unpolished part of the fiber to asymmetric on polished and back \cite{Nikbakht2019}.

\begin{figure*}
\includegraphics[width=0.8\linewidth,keepaspectratio]{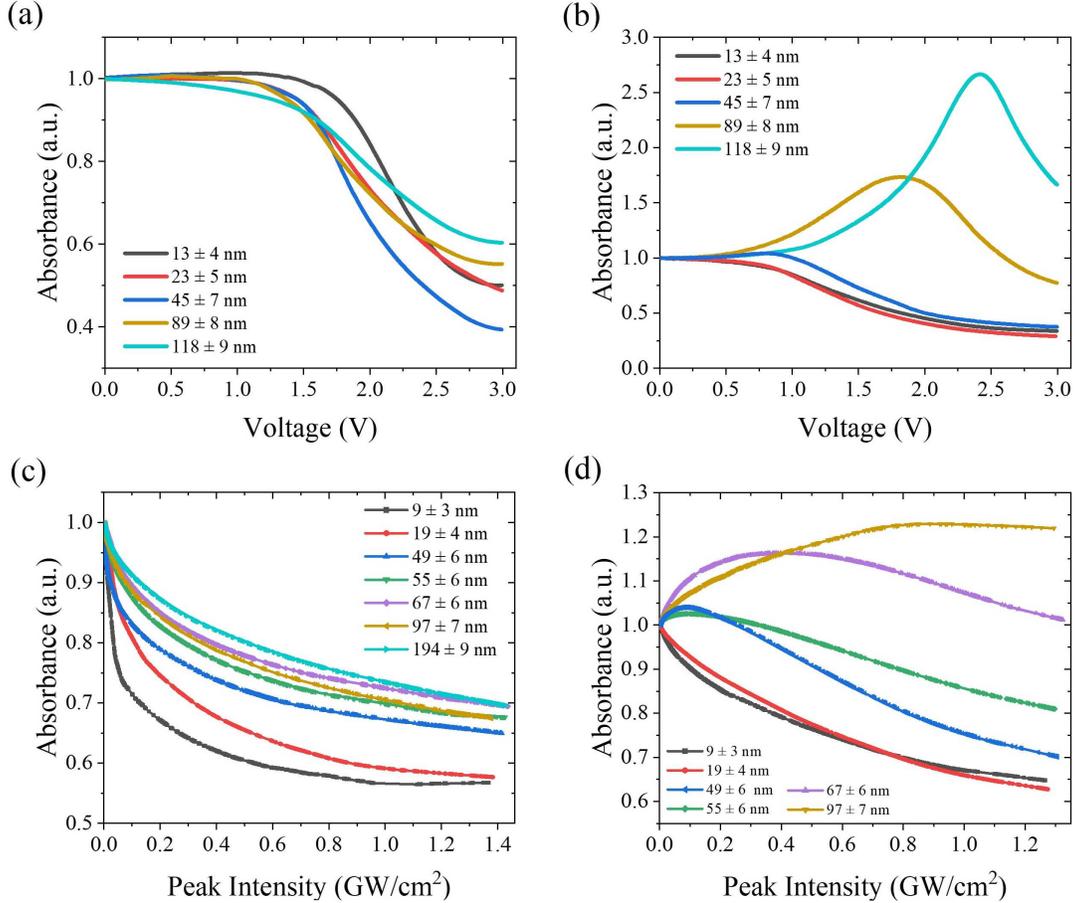}
\caption{\label{fig:fig2}Experimental absorbance through the waveguide at 1.55 $\mu$m in various configurations. (a) Absorbance curves of SWCNT films of different thicknesses (values are given in the caption) on a glass substrate during electrochemical gating normalized to the non-gated value. (b) Absorbance curves of the SPF covered with SWCNT films of different thicknesses during electrochemical gating normalized to the non-gated value. (c) Nonlinear absorbance measurements of SWCNTs film of different thicknesses on the fiber ferrule normalized to the small-signal value (when the SWCNT film is unsaturated). (d) Measured nonlinear absorbance of the SPF covered with SWCNT films of different thicknesses normalized to the small-signal value.}
\end{figure*}

Now we move to the experimental demonstration of the earlier discussed effect varying $\kappa_{\textrm {film}}$. For SWCNT films there are two possibilities to reduce the material loss coefficient: through the saturable absorption effect by the resonant excitation with ultrashort pulses \cite{Maeda2005,Xu2016} and by electrochemical gating \cite{Kazaoui2001}. We start with electrochemical gating, where the application of the potential to the SWCNT film immersed in the ionic liquid leads to an electrical double-layer formation that allows to accumulate large excess charge on the surface. It causes a shift of the Fermi level towards conduction or valence band and absorption reduction regardless of the voltage sign \cite{Bisri2017}. We perform our measurements in two configurations, when light propagates through the gated SWCNT film from free space (whole light propagates orthogonally to the carbon nanotube film) and through SPF-SWCNT. Note that we use CW laser for this measurement with power as small as 10 mW to avoid nonlinear effects and heating. For the free-space measurements in \autoref{fig:fig2}a, we observe that the film gets more transparent with the gating regardless of thickness. The absorbance dependence of SWCNT films on the SPF for TE polarization for various film thicknesses is shown in \autoref{fig:fig2}b. To compare it with numerical predictions in \autoref{fig:fig1}e all curves are normalized to non-gated value (non-normalized graphs can be seen in \hyperref[suppinfo]{Figure~S7 in Supplementary Information}). Here we see good qualitative agreement with numerically calculated results: 10 and 20 nm films demonstrate the decrease in absorbance, while for thicker films absorbance increases with applied voltage, reaches the maximum, and goes down. The difference in the curve shapes in \autoref{fig:fig1}e and \autoref{fig:fig2}b comes from an unknown functional dependence of material losses $\kappa_{\textrm {film}}$ of SWCNTs on the applied voltage.

Now we investigate how the mode reshaping affects saturable absorption, which is well-known in SWCNT under resonant excitation \cite{Lauret2003,Ostojic2004}. The nonlinear absorption is measured by a classic twin-detector method \cite{Gladush2019} using amplified signals from a homemade fully polarization-maintaining ultrafast fiber laser at 1.55 $\mu$m. First, we measure the nonlinear absorbance of the films deposited on the connectors, where the entire mode interacts with the film in the in-plane polarization, similar to free-space measurements with gating. For all the sample thicknesses we observe absorption saturation regardless of film thickness, see \autoref{fig:fig2}c. In case of SPF-SWCNT samples (see \autoref{fig:fig2}d), we observe thin films to become more transparent at high intensities while thick films demonstrate nonmonotonic behavior with increased absorption followed by absorption reduction as the intensity increases, in agreement with the numerical prediction in \autoref{fig:fig1}e We rule out a contribution of other possible effects to the induced absorption. The two-photon absorption or other reasons of optical limiting are not observed for resonant transitions for SWCNTs. Indeed, \autoref{fig:fig2}c evidences that the nonmonotonic behavior in \autoref{fig:fig2}d is not an intrinsic film property, but results from the geometry of light-matter interaction. To exclude thermal contribution to this behavior, we check that there is no nonlinear absorbance for the thickest SWCNT film on the SPF if a continuous wave laser with the same average power is used for nonlinear absorbance measurements (see \hyperref[suppinfo]{Figure~S10a in Supplementary Information}). Finally, we check that polarization rotation does not take place for SPF-SWCNT if the polarization of the light coincides on the entrance with slow or fast axis, see \hyperref[suppinfo]{Figure~S10b}.

Thus, regardless of how the material losses of SWCNTs are changed experimentally, the examined effect of the mode reshaping in the SPF with thick SWCNT films can be detected for the TE polarization. We do not observe the same effect for TM polarization which is also in agreement with numerical simulation. The results of the measurements and numerical model for out-of-plane polarization can be found in \hyperref[suppinfo]{Supplementary Information~S5}.

\subsection*{Analytical approach}

\begin{figure*}
\includegraphics[width=0.8\linewidth,keepaspectratio]{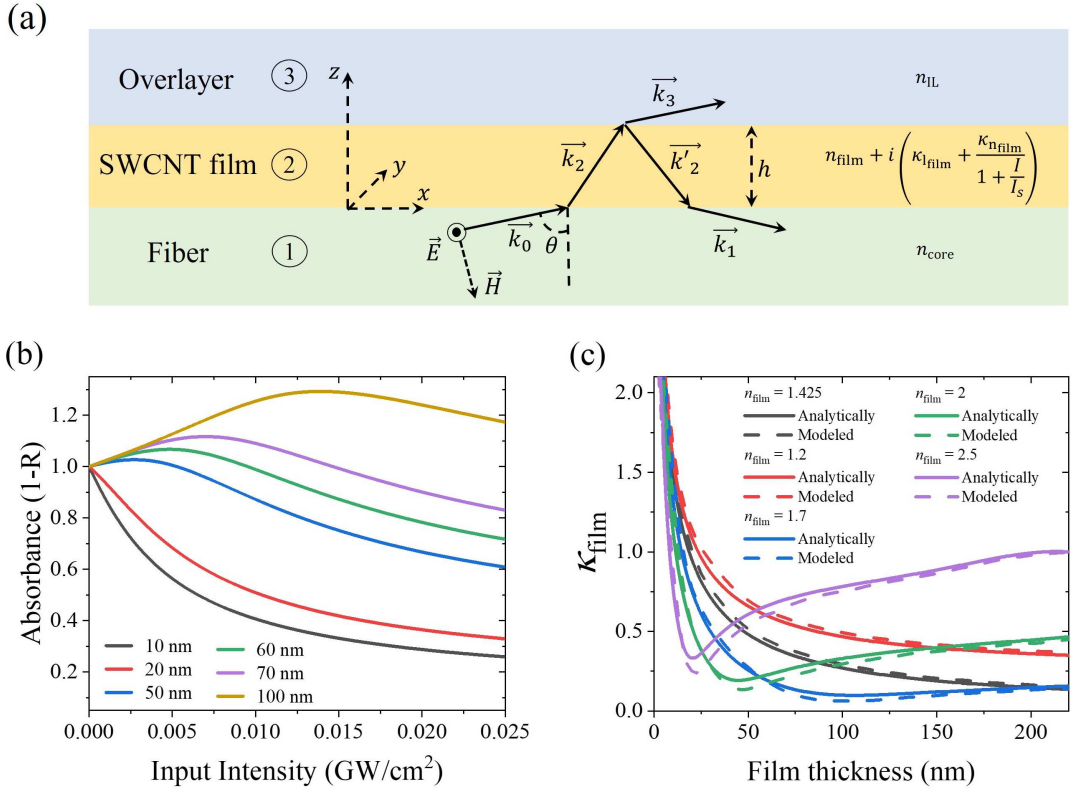}
\caption{\label{fig:fig3}(a) Sketch of virtual experiment for analytical approach: plane wave from media 1 incident under acute angle on the thin saturable absorber film (2) undergoes the total internal reflection from media 3. (b) Nonlinear absorbance, calculated as 1 – Reflection, from SWCNT films of different thicknesses normalized to the small-signal value. (c) The curves, defined by condition $(\frac{\mathrm{d}\alpha_\textrm{SPF}(n,d)}{\mathrm{d}\kappa_\textrm{film}} = 0)$, which divide the film parameter space in normal and reverse saturable absorption for different real parts of the SWCNT refractive index.}
\end{figure*}

The effect of switching saturable absorption to reverse effect can be resembled analytically if we consider a straightforward case of the plane wave reflected by the saturable absorber film under the total internal reflection condition. The scheme of the virtual experiment is shown in \autoref{fig:fig3}a. We consider a plane wave incident under an acute angle from an optically dense media to a thin layer with a saturable absorption followed by overlayer. To compare analysis with experimental results, we use the refractive indices of layers corresponding to the fiber core, carbon nanotubes, and ionic liquid (see \autoref{tab:table1}). We use an approach similar to Landau et al. \cite{Landau2013}, where we consider the intermediate layer's saturable absorption. The details of the calculations are shown in \hyperref[suppinfo]{Supplementary Information~S6}. \autoref{fig:fig3}a presents the dependences of the absorption for various SWCNT film thicknesses for the angle of incidence equal to 85°, which meets total internal reflection angle requirements of the optical fiber. The results are very similar to experimental observations of the nonlinear response of the SPF-SWCNT in \autoref{fig:fig2}d: absorption goes down with intensity for the small thickness while for 50 nm and higher absorption shows a gradual increase followed by absorption reduction. This effect preserves all the angles of wave incidence for given refractive indices if the total reflection condition is fulfilled. 

Finally, we would like to define a range of parameters of the film, where reverse saturable absorption can be observed. The regions with normal absorption and reverse effect are separated by the condition $(\frac{\mathrm{d}\alpha_\textrm{SPF}(n,d)}{\mathrm{d}\kappa_\textrm{film}} = 0)$, where absorption is maximum for a given film thickness and refractive index is reached. In \autoref{fig:fig3}c we plot curves satisfying this condition in coordinates of film thickness and $\kappa_{\textrm {film}}$ for various values of the real part refractive index of the film by using the analytical approach (solid lines) and data from \autoref{fig:fig1}d (dashed lines). Below the curve, a normal saturable absorption is expected, while above the curve a reversible effect should be observed. We find analytical curves and modeling results in remarkable agreement despite the differences in the modes of cylindrical waveguide and plane waves. The correspondence between the analytical approach and modeling is shown more explicitly in \hyperref[suppinfo]{Figure~S14b in Supplementary Information} for $n_{\textrm {film}}$ = 1.42. The largest deviations between theory and modeling are observed near the dips for the curves with larger values of the refractive index ($n_{\textrm {film}}$ = 1.7, 2, 2.5). These points correspond to the phase-matching condition of the modes in the fiber and in the film, leading to a situation where a significant part of the mode propagates within the film. This parameter regions cannot be handled properly by our simple analytical approach and mode propagation through coupled waveguides should be considered here.

\section*{Discussion}

The effect under consideration is general for waveguiding systems, covered by a thin absorbing film, including waveguides of the rectangular profile used in many applications from semiconductor lasers to integrated optics. It can also work in a reverse manner if the waveguide is covered by the material with optical limiting. In this case, the nonlinear increase of absorbance of covering material will turn into a saturable absorption for waveguide propagating light for a for a properly tuned film thickness and refractive index. It opens the perspectives for engineering of the nonlinear optical response which can be implemented in all-optical computing and neuromorphic integrated photonic systems \cite{Shastri2021}.

In summary, using carbon nanotube thin films deposited on a side-polished fiber, we have demonstrated that the light mode in the waveguide experiencing evanescent-field interaction with the covering material can be modified by changing the material losses of the covering film. It can lead to several practical effects like nonmonotonic dependence of waveguide losses on the overlaying film thickness and, more remarkably, turning of the saturable absorption in photoinduced absorption. We have shown that this effect is strongly dependent on the thickness of the film and can be observed if the thickness exceeds a certain value. Our analytical findings suggest that this effect is not limited to a side polished fiber configuration but can be observed in a wide range of waveguided systems covered by the absorbing material once requirements on refractive index and covering film thickness are met. 

\section*{Methods} \label{methods}

\subsection*{SWCNT film synthesis and transfer onto the SPF}
Single-walled carbon nanotube synthesis method employs aerosol CVD based on the Boudouard reaction with ferrocene acting as a catalyst precursor and CO as a carbon source described elsewhere \cite{Krasnikov2023}. A flow of carbon monoxide CO and Fe-based nanoparticles heated to 880° C in a tubular furnace reactor resulted in the formation of SWCNTs \cite{Iakovlev2020}. The films of the SWCNTs are collected downstream of the reactor on a nitrocellulose filter. The film is a mixture of semiconducting and metallic SWCNTs that form a random network with a predominant orientation in the film surface. The mean diameter of SWCNTs can be controlled during the synthesis process in the range from 1.0 to 2.3 nm by introducing a certain amount of carbon dioxide (C$\textrm{O}_2$) in the reactor \cite{Nasibulin2006,Khabushev2020}. The thickness of the SWCNT film on a filter can be controlled with high precision by the collection time. We use a commercial side-polished fiber (KS Photonics inc). SWCNT films are transferred onto the SPF by the dry-transfer technique: collected on the nitrocellulose filter, SWCNT films are pressed towards the polished fiber surface. Due to a weak adhesion to the filter and the high specific surface area of the SWCNTs, films can be easily dry transferred to practically any other substrates by simply pressing it down \cite{Mkrtchyan2019}. For the saturable absorption experiments the length of the film was chosen so that the small signal (unsaturated) loss through the fiber was around 3 dB to avoid effects of detector nonlinearity and significant intensity reduction along the covered part. For the densification, the SWCNT films were first soaked in a drop of ethanol and subsequently dried out. To fabricate an ionic cell, we used the same design as in previous work \cite{Gladush2019}, so another SWCNT film was transferred in the immediate vicinity for a counter electrode that does not cover the polished part and thus is not affecting the losses. The electrode wires are fixed to the block with a silver paste. Then, the ionic liquid (Bmim NT$\textrm{f}_2$) is dripped on the sample to cover both electrodes. Ionic liquid was synthesized by the standard procedure including alkylation 1-methylimidazole by 1-bromobutane and following anion exchange with Lithium bistriflimide in the water phase \cite{Vorotyntsev2009}.

\subsection*{SWCNT films and ionic liquid constants measurements}

The complex refractive indices of the SWCNT film and ionic liquid are measured at 1.55 $\mu$m. The imaginary part of the refractive index of SWCNTs is retrieved from fitting absorbance of nanotube films of different thicknesses deposited on fiber ferrules: $\kappa_\textrm{film}$ = 0.62±0.06 (\hyperref[suppinfo]{Figure~S3 in Supplementary Information}). We measure the real part of the refractive index by measuring the reflected power from the fiber end with a deposited SWCNT film thick enough to absorb all the light so reflection from the second interface can be neglected. Similarly, the refractive index of ionic liquid was measured by dripping fiber end into the liquid (see \hyperref[suppinfo]{Figure~S4}). The refractive index for carbon nanotubes, calculated using Fresnel equation and known refractive index of the fiber core, varies between 1.35 and 1.6 and were chosen as $n_\textrm{film}$ = 1.43 to provide best correspondence with experimental results. A large range of its values is attributed to a variation in nanotube density during liquid densification and transfer onto the fiber ferrule. The refractive index of the ionic liquid is $n_\textrm{IL}$ = 1.41±0.01 in good agreement with the extrapolated values from the literature \cite{Montalban2015, Wu2018,Arosa2018}. As a reference we measure the refractive index of water: $n_\textrm{water}$ = 1.319. The deviation from the known value ($n_\textrm{water}$ = 1.318 \cite{Hale1973}) is in the third decimal place which confirms the high accuracy of the method.

\subsection*{Eletrochemical gating and nonlinear transmittance measurements}
As a light source for measurements with electrochemical gating, we use Pure Photonics PPCL550 tunable laser at 1.55 $\mu$m with narrow 10kHz linewidth and ultra-low FM noise. In the case of free-space measurements, an ionic cell with two SWCNT film electrodes on a glass substrate is prepared using SWCNT dry transfer technique. We utilize Potentiostat R-45X with module for measuring electrochemical impedance FRA-24M by Electro Chemical Instruments to apply voltage and monitor electrochemical characteristics of gated SWCNTs. Light power is monitored using the power meter Thorlabs PM100D. 

For nonlinear optical absorption measurements, we use a homemade fully polarization-maintaining ultrafast fiber laser at 1.55 $\mu$m as a source of pulses, which is mode-locked by the SWCNT film on the fiber ferrules. The commercial erbium-doped fiber amplifier Keopsys PEFA-SP-C-PM-27-B130-FA-FA is employed to amplify the signal from the laser. Then, the light is being attenuated and splitted by a high-power coupler. One output of the coupler works as a reference and another one is used for exposure of the sample. We monitor the light power using the dual-channel optical power meter Thorlabs PM320E. The scheme of the described experimental setup can be seen in \hyperref[suppinfo]{Figure~S8 in Supplementary Information}. Measured pulse autocorrelation and optical spectrum can be found in \hyperref[suppinfo]{Figure~S9}.

\bibliography{main}

\begin{thebibliography}{65}%
\makeatletter
\providecommand \@ifxundefined [1]{%
 \@ifx{#1\undefined}
}%
\providecommand \@ifnum [1]{%
 \ifnum #1\expandafter \@firstoftwo
 \else \expandafter \@secondoftwo
 \fi
}%
\providecommand \@ifx [1]{%
 \ifx #1\expandafter \@firstoftwo
 \else \expandafter \@secondoftwo
 \fi
}%
\providecommand \natexlab [1]{#1}%
\providecommand \enquote  [1]{``#1''}%
\providecommand \bibnamefont  [1]{#1}%
\providecommand \bibfnamefont [1]{#1}%
\providecommand \citenamefont [1]{#1}%
\providecommand \href@noop [0]{\@secondoftwo}%
\providecommand \href [0]{\begingroup \@sanitize@url \@href}%
\providecommand \@href[1]{\@@startlink{#1}\@@href}%
\providecommand \@@href[1]{\endgroup#1\@@endlink}%
\providecommand \@sanitize@url [0]{\catcode `\\12\catcode `\$12\catcode
  `\&12\catcode `\#12\catcode `\^12\catcode `\_12\catcode `\%12\relax}%
\providecommand \@@startlink[1]{}%
\providecommand \@@endlink[0]{}%
\providecommand \url  [0]{\begingroup\@sanitize@url \@url }%
\providecommand \@url [1]{\endgroup\@href {#1}{\urlprefix }}%
\providecommand \urlprefix  [0]{URL }%
\providecommand \Eprint [0]{\href }%
\providecommand \doibase [0]{https://doi.org/}%
\providecommand \selectlanguage [0]{\@gobble}%
\providecommand \bibinfo  [0]{\@secondoftwo}%
\providecommand \bibfield  [0]{\@secondoftwo}%
\providecommand \translation [1]{[#1]}%
\providecommand \BibitemOpen [0]{}%
\providecommand \bibitemStop [0]{}%
\providecommand \bibitemNoStop [0]{.\EOS\space}%
\providecommand \EOS [0]{\spacefactor3000\relax}%
\providecommand \BibitemShut  [1]{\csname bibitem#1\endcsname}%
\let\auto@bib@innerbib\@empty
\bibitem [{\citenamefont {Romagnoli}\ \emph {et~al.}(2018)\citenamefont
  {Romagnoli}, \citenamefont {Sorianello}, \citenamefont {Midrio},
  \citenamefont {Koppens}, \citenamefont {Huyghebaert}, \citenamefont
  {Neumaier}, \citenamefont {Galli}, \citenamefont {Templ}, \citenamefont
  {D'Errico},\ and\ \citenamefont {Ferrari}}]{Romagnoli2018}%
  \BibitemOpen
  \bibfield  {author} {\bibinfo {author} {\bibfnamefont {M.}~\bibnamefont
  {Romagnoli}}, \bibinfo {author} {\bibfnamefont {V.}~\bibnamefont
  {Sorianello}}, \bibinfo {author} {\bibfnamefont {M.}~\bibnamefont {Midrio}},
  \bibinfo {author} {\bibfnamefont {F.~H.}\ \bibnamefont {Koppens}}, \bibinfo
  {author} {\bibfnamefont {C.}~\bibnamefont {Huyghebaert}}, \bibinfo {author}
  {\bibfnamefont {D.}~\bibnamefont {Neumaier}}, \bibinfo {author}
  {\bibfnamefont {P.}~\bibnamefont {Galli}}, \bibinfo {author} {\bibfnamefont
  {W.}~\bibnamefont {Templ}}, \bibinfo {author} {\bibfnamefont
  {A.}~\bibnamefont {D'Errico}},\ and\ \bibinfo {author} {\bibfnamefont
  {A.~C.}\ \bibnamefont {Ferrari}},\ }\bibfield  {title} {\bibinfo {title}
  {{Graphene-based integrated photonics for next-generation datacom and
  telecom}},\ }\href {https://doi.org/10.1038/s41578-018-0040-9} {\bibfield
  {journal} {\bibinfo  {journal} {Nature Reviews Materials}\ }\textbf {\bibinfo
  {volume} {3}},\ \bibinfo {pages} {392} (\bibinfo {year} {2018})}\BibitemShut
  {NoStop}%
\bibitem [{\citenamefont {Youngblood}\ \emph {et~al.}(2014)\citenamefont
  {Youngblood}, \citenamefont {Anugrah}, \citenamefont {Ma}, \citenamefont
  {Koester},\ and\ \citenamefont {Li}}]{Youngblood2014}%
  \BibitemOpen
  \bibfield  {author} {\bibinfo {author} {\bibfnamefont {N.}~\bibnamefont
  {Youngblood}}, \bibinfo {author} {\bibfnamefont {Y.}~\bibnamefont {Anugrah}},
  \bibinfo {author} {\bibfnamefont {R.}~\bibnamefont {Ma}}, \bibinfo {author}
  {\bibfnamefont {S.~J.}\ \bibnamefont {Koester}},\ and\ \bibinfo {author}
  {\bibfnamefont {M.}~\bibnamefont {Li}},\ }\bibfield  {title} {\bibinfo
  {title} {{Multifunctional graphene optical modulator and photodetector
  integrated on silicon waveguides}},\ }\href
  {https://doi.org/10.1021/NL500712U/SUPPL_FILE/NL500712U_SI_001.PDF}
  {\bibfield  {journal} {\bibinfo  {journal} {Nano Letters}\ }\textbf {\bibinfo
  {volume} {14}},\ \bibinfo {pages} {2741} (\bibinfo {year}
  {2014})}\BibitemShut {NoStop}%
\bibitem [{\citenamefont {Bonaccorso}\ \emph {et~al.}(2010)\citenamefont
  {Bonaccorso}, \citenamefont {Sun}, \citenamefont {Hasan},\ and\ \citenamefont
  {Ferrari}}]{Bonaccorso2010}%
  \BibitemOpen
  \bibfield  {author} {\bibinfo {author} {\bibfnamefont {F.}~\bibnamefont
  {Bonaccorso}}, \bibinfo {author} {\bibfnamefont {Z.}~\bibnamefont {Sun}},
  \bibinfo {author} {\bibfnamefont {T.}~\bibnamefont {Hasan}},\ and\ \bibinfo
  {author} {\bibfnamefont {A.~C.}\ \bibnamefont {Ferrari}},\ }\bibfield
  {title} {\bibinfo {title} {{Graphene photonics and optoelectronics}},\ }\href
  {https://doi.org/10.1038/nphoton.2010.186} {\bibfield  {journal} {\bibinfo
  {journal} {Nature Photonics}\ }\textbf {\bibinfo {volume} {4}},\ \bibinfo
  {pages} {611} (\bibinfo {year} {2010})}\BibitemShut {NoStop}%
\bibitem [{\citenamefont {Bao}\ and\ \citenamefont {Loh}(2012)}]{Bao2012}%
  \BibitemOpen
  \bibfield  {author} {\bibinfo {author} {\bibfnamefont {Q.}~\bibnamefont
  {Bao}}\ and\ \bibinfo {author} {\bibfnamefont {K.~P.}\ \bibnamefont {Loh}},\
  }\bibfield  {title} {\bibinfo {title} {{Graphene photonics, plasmonics, and
  broadband optoelectronic devices}},\ }\href
  {https://doi.org/10.1021/NN300989G} {\bibfield  {journal} {\bibinfo
  {journal} {ACS Nano}\ }\textbf {\bibinfo {volume} {6}},\ \bibinfo {pages}
  {3677} (\bibinfo {year} {2012})}\BibitemShut {NoStop}%
\bibitem [{\citenamefont {Jeong}\ \emph {et~al.}(2013)\citenamefont {Jeong},
  \citenamefont {Choi}, \citenamefont {Jeong}, \citenamefont {Cha},
  \citenamefont {Rotermund},\ and\ \citenamefont {Yeom}}]{Jeong2013}%
  \BibitemOpen
  \bibfield  {author} {\bibinfo {author} {\bibfnamefont {H.}~\bibnamefont
  {Jeong}}, \bibinfo {author} {\bibfnamefont {S.~Y.}\ \bibnamefont {Choi}},
  \bibinfo {author} {\bibfnamefont {E.~I.}\ \bibnamefont {Jeong}}, \bibinfo
  {author} {\bibfnamefont {S.~J.}\ \bibnamefont {Cha}}, \bibinfo {author}
  {\bibfnamefont {F.}~\bibnamefont {Rotermund}},\ and\ \bibinfo {author}
  {\bibfnamefont {D.-I.}\ \bibnamefont {Yeom}},\ }\bibfield  {title} {\bibinfo
  {title} {{Ultrafast Mode-Locked Fiber Laser Using a Waveguide-Type Saturable
  Absorber Based on Single-Walled Carbon Nanotubes}},\ }\href
  {https://doi.org/10.7567/APEX.6.052705} {\bibfield  {journal} {\bibinfo
  {journal} {Applied Physics Express}\ }\textbf {\bibinfo {volume} {6}},\
  \bibinfo {pages} {052705} (\bibinfo {year} {2013})}\BibitemShut {NoStop}%
\bibitem [{\citenamefont {Tan}\ \emph {et~al.}(2020)\citenamefont {Tan},
  \citenamefont {Jiang}, \citenamefont {Wang}, \citenamefont {Yao},
  \citenamefont {Zhang}, \citenamefont {Tan}, \citenamefont {Yao},
  \citenamefont {Jiang}, \citenamefont {Wang},\ and\ \citenamefont
  {Zhang}}]{Tan2020}%
  \BibitemOpen
  \bibfield  {author} {\bibinfo {author} {\bibfnamefont {T.}~\bibnamefont
  {Tan}}, \bibinfo {author} {\bibfnamefont {X.}~\bibnamefont {Jiang}}, \bibinfo
  {author} {\bibfnamefont {C.}~\bibnamefont {Wang}}, \bibinfo {author}
  {\bibfnamefont {B.}~\bibnamefont {Yao}}, \bibinfo {author} {\bibfnamefont
  {H.}~\bibnamefont {Zhang}}, \bibinfo {author} {\bibfnamefont
  {T.}~\bibnamefont {Tan}}, \bibinfo {author} {\bibfnamefont {B.~C.}\
  \bibnamefont {Yao}}, \bibinfo {author} {\bibfnamefont {X.~T.}\ \bibnamefont
  {Jiang}}, \bibinfo {author} {\bibfnamefont {C.}~\bibnamefont {Wang}},\ and\
  \bibinfo {author} {\bibfnamefont {H.}~\bibnamefont {Zhang}},\ }\bibfield
  {title} {\bibinfo {title} {{2D Material Optoelectronics for Information
  Functional Device Applications: Status and Challenges}},\ }\href
  {https://doi.org/10.1002/ADVS.202000058} {\bibfield  {journal} {\bibinfo
  {journal} {Advanced Science}\ }\textbf {\bibinfo {volume} {7}},\ \bibinfo
  {pages} {2000058} (\bibinfo {year} {2020})}\BibitemShut {NoStop}%
\bibitem [{\citenamefont {Sun}\ \emph {et~al.}(2016)\citenamefont {Sun},
  \citenamefont {Martinez},\ and\ \citenamefont {Wang}}]{Sun2016}%
  \BibitemOpen
  \bibfield  {author} {\bibinfo {author} {\bibfnamefont {Z.}~\bibnamefont
  {Sun}}, \bibinfo {author} {\bibfnamefont {A.}~\bibnamefont {Martinez}},\ and\
  \bibinfo {author} {\bibfnamefont {F.}~\bibnamefont {Wang}},\ }\bibfield
  {title} {\bibinfo {title} {{Optical modulators with 2D layered materials}},\
  }\href {https://doi.org/10.1038/nphoton.2016.15} {\bibfield  {journal}
  {\bibinfo  {journal} {Nature Photonics}\ }\textbf {\bibinfo {volume} {10}},\
  \bibinfo {pages} {227} (\bibinfo {year} {2016})}\BibitemShut {NoStop}%
\bibitem [{\citenamefont {Yu}\ \emph {et~al.}(2017)\citenamefont {Yu},
  \citenamefont {Wu}, \citenamefont {Wang}, \citenamefont {Guo},\ and\
  \citenamefont {Tong}}]{Yu2017}%
  \BibitemOpen
  \bibfield  {author} {\bibinfo {author} {\bibfnamefont {S.}~\bibnamefont
  {Yu}}, \bibinfo {author} {\bibfnamefont {X.}~\bibnamefont {Wu}}, \bibinfo
  {author} {\bibfnamefont {Y.}~\bibnamefont {Wang}}, \bibinfo {author}
  {\bibfnamefont {X.}~\bibnamefont {Guo}},\ and\ \bibinfo {author}
  {\bibfnamefont {L.}~\bibnamefont {Tong}},\ }\bibfield  {title} {\bibinfo
  {title} {{2D Materials for Optical Modulation: Challenges and
  Opportunities}},\ }\href {https://doi.org/10.1002/ADMA.201606128} {\bibfield
  {journal} {\bibinfo  {journal} {Advanced Materials}\ }\textbf {\bibinfo
  {volume} {29}},\ \bibinfo {pages} {1606128} (\bibinfo {year}
  {2017})}\BibitemShut {NoStop}%
\bibitem [{\citenamefont {Parra}\ \emph {et~al.}(2021)\citenamefont {Parra},
  \citenamefont {Pernice},\ and\ \citenamefont {Sanchis}}]{Parra2021}%
  \BibitemOpen
  \bibfield  {author} {\bibinfo {author} {\bibfnamefont {J.}~\bibnamefont
  {Parra}}, \bibinfo {author} {\bibfnamefont {W.~H.}\ \bibnamefont {Pernice}},\
  and\ \bibinfo {author} {\bibfnamefont {P.}~\bibnamefont {Sanchis}},\
  }\bibfield  {title} {\bibinfo {title} {{All-optical phase control in
  nanophotonic silicon waveguides with epsilon-near-zero nanoheaters}},\ }\href
  {https://doi.org/10.1038/s41598-021-88865-6} {\bibfield  {journal} {\bibinfo
  {journal} {Scientific Reports}\ }\textbf {\bibinfo {volume} {11}},\ \bibinfo
  {pages} {1} (\bibinfo {year} {2021})}\BibitemShut {NoStop}%
\bibitem [{\citenamefont {Janjan}\ \emph {et~al.}(2017)\citenamefont {Janjan},
  \citenamefont {Miri}, \citenamefont {Heidari},\ and\ \citenamefont
  {Zarifkar}}]{Janjan2017}%
  \BibitemOpen
  \bibfield  {author} {\bibinfo {author} {\bibfnamefont {B.}~\bibnamefont
  {Janjan}}, \bibinfo {author} {\bibfnamefont {M.}~\bibnamefont {Miri}},
  \bibinfo {author} {\bibfnamefont {M.}~\bibnamefont {Heidari}},\ and\ \bibinfo
  {author} {\bibfnamefont {A.}~\bibnamefont {Zarifkar}},\ }\bibfield  {title}
  {\bibinfo {title} {{Design and Simulation of Compact Optical Modulators and
  Switches Based on Si–VO2–Si Horizontal Slot Waveguides}},\ }\href
  {https://opg.optica.org/abstract.cfm?uri=jlt-35-14-3020} {\bibfield
  {journal} {\bibinfo  {journal} {Journal of Lightwave Technology}\ }\textbf
  {\bibinfo {volume} {35}},\ \bibinfo {pages} {3020} (\bibinfo {year}
  {2017})}\BibitemShut {NoStop}%
\bibitem [{\citenamefont {Khan}\ and\ \citenamefont {Kang}(2014)}]{Khan2014}%
  \BibitemOpen
  \bibfield  {author} {\bibinfo {author} {\bibfnamefont {M.~R.~R.}\
  \bibnamefont {Khan}}\ and\ \bibinfo {author} {\bibfnamefont {S.~W.}\
  \bibnamefont {Kang}},\ }\bibfield  {title} {\bibinfo {title} {{A high
  sensitivity and wide dynamic range fiber-optic sensor for low-concentration
  VOC gas detection}},\ }\href {https://doi.org/10.3390/s141223321} {\bibfield
  {journal} {\bibinfo  {journal} {Sensor}\ }\textbf {\bibinfo {volume} {14}},\
  \bibinfo {pages} {23321} (\bibinfo {year} {2014})}\BibitemShut {NoStop}%
\bibitem [{\citenamefont {Hu}\ \emph {et~al.}(2018)\citenamefont {Hu},
  \citenamefont {Luo}, \citenamefont {Guan}, \citenamefont {Zhu}, \citenamefont
  {Yu}, \citenamefont {Chen}, \citenamefont {Chen}, \citenamefont {Dong},
  \citenamefont {Jiang}, \citenamefont {Zhang} \emph {et~al.}}]{Hu2018}%
  \BibitemOpen
  \bibfield  {author} {\bibinfo {author} {\bibfnamefont {S.}~\bibnamefont
  {Hu}}, \bibinfo {author} {\bibfnamefont {Y.}~\bibnamefont {Luo}}, \bibinfo
  {author} {\bibfnamefont {H.}~\bibnamefont {Guan}}, \bibinfo {author}
  {\bibfnamefont {W.}~\bibnamefont {Zhu}}, \bibinfo {author} {\bibfnamefont
  {J.}~\bibnamefont {Yu}}, \bibinfo {author} {\bibfnamefont {Z.}~\bibnamefont
  {Chen}}, \bibinfo {author} {\bibfnamefont {Y.}~\bibnamefont {Chen}}, \bibinfo
  {author} {\bibfnamefont {J.}~\bibnamefont {Dong}}, \bibinfo {author}
  {\bibfnamefont {Z.}~\bibnamefont {Jiang}}, \bibinfo {author} {\bibfnamefont
  {J.}~\bibnamefont {Zhang}}, \emph {et~al.},\ }\bibfield  {title} {\bibinfo
  {title} {{High-sensitivity vector magnetic field sensor based on
  side-polished fiber plasmon and ferrofluid}},\ }\href
  {https://doi.org/10.1364/OL.43.004743} {\bibfield  {journal} {\bibinfo
  {journal} {Optics Letters}\ }\textbf {\bibinfo {volume} {43}},\ \bibinfo
  {pages} {4743} (\bibinfo {year} {2018})}\BibitemShut {NoStop}%
\bibitem [{\citenamefont {An}\ \emph {et~al.}(2020)\citenamefont {An},
  \citenamefont {Tan}, \citenamefont {Peng}, \citenamefont {Qin}, \citenamefont
  {Yuan}, \citenamefont {Bi}, \citenamefont {Liao}, \citenamefont {Wang},
  \citenamefont {Rao}, \citenamefont {Soavi} \emph {et~al.}}]{An2020}%
  \BibitemOpen
  \bibfield  {author} {\bibinfo {author} {\bibfnamefont {N.}~\bibnamefont
  {An}}, \bibinfo {author} {\bibfnamefont {T.}~\bibnamefont {Tan}}, \bibinfo
  {author} {\bibfnamefont {Z.}~\bibnamefont {Peng}}, \bibinfo {author}
  {\bibfnamefont {C.}~\bibnamefont {Qin}}, \bibinfo {author} {\bibfnamefont
  {Z.}~\bibnamefont {Yuan}}, \bibinfo {author} {\bibfnamefont {L.}~\bibnamefont
  {Bi}}, \bibinfo {author} {\bibfnamefont {C.}~\bibnamefont {Liao}}, \bibinfo
  {author} {\bibfnamefont {Y.}~\bibnamefont {Wang}}, \bibinfo {author}
  {\bibfnamefont {Y.}~\bibnamefont {Rao}}, \bibinfo {author} {\bibfnamefont
  {G.}~\bibnamefont {Soavi}}, \emph {et~al.},\ }\bibfield  {title} {\bibinfo
  {title} {{Electrically Tunable Four-Wave-Mixing in Graphene Heterogeneous
  Fiber for Individual Gas Molecule Detection}},\ }\href
  {https://doi.org/10.1021/ACS.NANOLETT.0C02174/ASSET/IMAGES/LARGE/NL0C02174_0004.JPEG}
  {\bibfield  {journal} {\bibinfo  {journal} {Nano Letters}\ }\textbf {\bibinfo
  {volume} {20}},\ \bibinfo {pages} {6473} (\bibinfo {year}
  {2020})}\BibitemShut {NoStop}%
\bibitem [{\citenamefont {Cao}\ \emph {et~al.}(2019)\citenamefont {Cao},
  \citenamefont {Yao}, \citenamefont {Qin}, \citenamefont {Yang}, \citenamefont
  {Guo}, \citenamefont {Zhang}, \citenamefont {Wu}, \citenamefont {Bi},
  \citenamefont {Chen}, \citenamefont {Xie} \emph {et~al.}}]{Cao2019}%
  \BibitemOpen
  \bibfield  {author} {\bibinfo {author} {\bibfnamefont {Z.}~\bibnamefont
  {Cao}}, \bibinfo {author} {\bibfnamefont {B.}~\bibnamefont {Yao}}, \bibinfo
  {author} {\bibfnamefont {C.}~\bibnamefont {Qin}}, \bibinfo {author}
  {\bibfnamefont {R.}~\bibnamefont {Yang}}, \bibinfo {author} {\bibfnamefont
  {Y.}~\bibnamefont {Guo}}, \bibinfo {author} {\bibfnamefont {Y.}~\bibnamefont
  {Zhang}}, \bibinfo {author} {\bibfnamefont {Y.}~\bibnamefont {Wu}}, \bibinfo
  {author} {\bibfnamefont {L.}~\bibnamefont {Bi}}, \bibinfo {author}
  {\bibfnamefont {Y.}~\bibnamefont {Chen}}, \bibinfo {author} {\bibfnamefont
  {Z.}~\bibnamefont {Xie}}, \emph {et~al.},\ }\bibfield  {title} {\bibinfo
  {title} {{Biochemical sensing in graphene-enhanced microfiber resonators with
  individual molecule sensitivity and selectivity}},\ }\href
  {https://doi.org/10.1038/s41377-019-0213-3} {\bibfield  {journal} {\bibinfo
  {journal} {Light: Science \& Applications}\ }\textbf {\bibinfo {volume}
  {8}},\ \bibinfo {pages} {1} (\bibinfo {year} {2019})}\BibitemShut {NoStop}%
\bibitem [{\citenamefont {Yan}\ \emph {et~al.}(2015)\citenamefont {Yan},
  \citenamefont {Zheng}, \citenamefont {Chen}, \citenamefont {Xu},\ and\
  \citenamefont {Lu}}]{Yan2015}%
  \BibitemOpen
  \bibfield  {author} {\bibinfo {author} {\bibfnamefont {S.~C.}\ \bibnamefont
  {Yan}}, \bibinfo {author} {\bibfnamefont {B.~C.}\ \bibnamefont {Zheng}},
  \bibinfo {author} {\bibfnamefont {J.~H.}\ \bibnamefont {Chen}}, \bibinfo
  {author} {\bibfnamefont {F.}~\bibnamefont {Xu}},\ and\ \bibinfo {author}
  {\bibfnamefont {Y.~Q.}\ \bibnamefont {Lu}},\ }\bibfield  {title} {\bibinfo
  {title} {{Optical electrical current sensor utilizing a
  graphene-microfiber-integrated coil resonator}},\ }\href
  {https://doi.org/10.1063/1.4928247} {\bibfield  {journal} {\bibinfo
  {journal} {Applied Physics Letters}\ }\textbf {\bibinfo {volume} {107}},\
  \bibinfo {pages} {053502} (\bibinfo {year} {2015})}\BibitemShut {NoStop}%
\bibitem [{\citenamefont {Zheng}\ \emph {et~al.}(2015)\citenamefont {Zheng},
  \citenamefont {Yan}, \citenamefont {Chen}, \citenamefont {Cui}, \citenamefont
  {Xu},\ and\ \citenamefont {Lu}}]{Zheng2015}%
  \BibitemOpen
  \bibfield  {author} {\bibinfo {author} {\bibfnamefont {B.~C.}\ \bibnamefont
  {Zheng}}, \bibinfo {author} {\bibfnamefont {S.~C.}\ \bibnamefont {Yan}},
  \bibinfo {author} {\bibfnamefont {J.~H.}\ \bibnamefont {Chen}}, \bibinfo
  {author} {\bibfnamefont {G.~X.}\ \bibnamefont {Cui}}, \bibinfo {author}
  {\bibfnamefont {F.}~\bibnamefont {Xu}},\ and\ \bibinfo {author}
  {\bibfnamefont {Y.~Q.}\ \bibnamefont {Lu}},\ }\bibfield  {title} {\bibinfo
  {title} {{Miniature optical fiber current sensor based on a graphene
  membrane}},\ }\href {https://doi.org/10.1002/LPOR.201500077} {\bibfield
  {journal} {\bibinfo  {journal} {Laser \& Photonics Reviews}\ }\textbf
  {\bibinfo {volume} {9}},\ \bibinfo {pages} {517} (\bibinfo {year}
  {2015})}\BibitemShut {NoStop}%
\bibitem [{\citenamefont {Abouraddy}\ \emph {et~al.}(2007)\citenamefont
  {Abouraddy}, \citenamefont {Bayindir}, \citenamefont {Benoit}, \citenamefont
  {Hart}, \citenamefont {Kuriki}, \citenamefont {Orf}, \citenamefont {Shapira},
  \citenamefont {Sorin}, \citenamefont {Temelkuran},\ and\ \citenamefont
  {Fink}}]{Abouraddy2007}%
  \BibitemOpen
  \bibfield  {author} {\bibinfo {author} {\bibfnamefont {A.~F.}\ \bibnamefont
  {Abouraddy}}, \bibinfo {author} {\bibfnamefont {M.}~\bibnamefont {Bayindir}},
  \bibinfo {author} {\bibfnamefont {G.}~\bibnamefont {Benoit}}, \bibinfo
  {author} {\bibfnamefont {S.~D.}\ \bibnamefont {Hart}}, \bibinfo {author}
  {\bibfnamefont {K.}~\bibnamefont {Kuriki}}, \bibinfo {author} {\bibfnamefont
  {N.}~\bibnamefont {Orf}}, \bibinfo {author} {\bibfnamefont {O.}~\bibnamefont
  {Shapira}}, \bibinfo {author} {\bibfnamefont {F.}~\bibnamefont {Sorin}},
  \bibinfo {author} {\bibfnamefont {B.}~\bibnamefont {Temelkuran}},\ and\
  \bibinfo {author} {\bibfnamefont {Y.}~\bibnamefont {Fink}},\ }\bibfield
  {title} {\bibinfo {title} {{Towards multimaterial multifunctional fibres that
  see, hear, sense and communicate}},\ }\href
  {https://doi.org/10.1038/nmat1889} {\bibfield  {journal} {\bibinfo  {journal}
  {Nature Materials}\ }\textbf {\bibinfo {volume} {6}},\ \bibinfo {pages} {336}
  (\bibinfo {year} {2007})}\BibitemShut {NoStop}%
\bibitem [{\citenamefont {Wang}\ \emph {et~al.}(2015)\citenamefont {Wang},
  \citenamefont {Zhao}, \citenamefont {Han}, \citenamefont {Fang},
  \citenamefont {Mao}, \citenamefont {Gan},\ and\ \citenamefont
  {Zhao}}]{Wang2015}%
  \BibitemOpen
  \bibfield  {author} {\bibinfo {author} {\bibfnamefont {Y.}~\bibnamefont
  {Wang}}, \bibinfo {author} {\bibfnamefont {C.}~\bibnamefont {Zhao}}, \bibinfo
  {author} {\bibfnamefont {L.}~\bibnamefont {Han}}, \bibinfo {author}
  {\bibfnamefont {L.}~\bibnamefont {Fang}}, \bibinfo {author} {\bibfnamefont
  {D.}~\bibnamefont {Mao}}, \bibinfo {author} {\bibfnamefont {X.}~\bibnamefont
  {Gan}},\ and\ \bibinfo {author} {\bibfnamefont {J.}~\bibnamefont {Zhao}},\
  }\bibfield  {title} {\bibinfo {title} {{Graphene-assisted all-fiber phase
  shifter and switching}},\ }\href {https://doi.org/10.1364/OPTICA.2.000468}
  {\bibfield  {journal} {\bibinfo  {journal} {Optica}\ }\textbf {\bibinfo
  {volume} {2}},\ \bibinfo {pages} {468} (\bibinfo {year} {2015})}\BibitemShut
  {NoStop}%
\bibitem [{\citenamefont {Chen}\ \emph {et~al.}(2015)\citenamefont {Chen},
  \citenamefont {Zheng}, \citenamefont {Shao}, \citenamefont {Ge},
  \citenamefont {Xu},\ and\ \citenamefont {Lu}}]{Chen2015}%
  \BibitemOpen
  \bibfield  {author} {\bibinfo {author} {\bibfnamefont {J.~H.}\ \bibnamefont
  {Chen}}, \bibinfo {author} {\bibfnamefont {B.~C.}\ \bibnamefont {Zheng}},
  \bibinfo {author} {\bibfnamefont {G.~H.}\ \bibnamefont {Shao}}, \bibinfo
  {author} {\bibfnamefont {S.~J.}\ \bibnamefont {Ge}}, \bibinfo {author}
  {\bibfnamefont {F.}~\bibnamefont {Xu}},\ and\ \bibinfo {author}
  {\bibfnamefont {Y.~Q.}\ \bibnamefont {Lu}},\ }\bibfield  {title} {\bibinfo
  {title} {{An all-optical modulator based on a stereo graphene–microfiber
  structure}},\ }\href {https://doi.org/10.1038/lsa.2015.133} {\bibfield
  {journal} {\bibinfo  {journal} {Light: Science \& Applications}\ }\textbf
  {\bibinfo {volume} {4}},\ \bibinfo {pages} {e360} (\bibinfo {year}
  {2015})}\BibitemShut {NoStop}%
\bibitem [{\citenamefont {Li}\ \emph {et~al.}(2014)\citenamefont {Li},
  \citenamefont {Chen}, \citenamefont {Meng}, \citenamefont {Fang},
  \citenamefont {Xiao}, \citenamefont {Li}, \citenamefont {Hu}, \citenamefont
  {Xu}, \citenamefont {Tong}, \citenamefont {Wang} \emph {et~al.}}]{Li2014}%
  \BibitemOpen
  \bibfield  {author} {\bibinfo {author} {\bibfnamefont {W.}~\bibnamefont
  {Li}}, \bibinfo {author} {\bibfnamefont {B.}~\bibnamefont {Chen}}, \bibinfo
  {author} {\bibfnamefont {C.}~\bibnamefont {Meng}}, \bibinfo {author}
  {\bibfnamefont {W.}~\bibnamefont {Fang}}, \bibinfo {author} {\bibfnamefont
  {Y.}~\bibnamefont {Xiao}}, \bibinfo {author} {\bibfnamefont {X.}~\bibnamefont
  {Li}}, \bibinfo {author} {\bibfnamefont {Z.}~\bibnamefont {Hu}}, \bibinfo
  {author} {\bibfnamefont {Y.}~\bibnamefont {Xu}}, \bibinfo {author}
  {\bibfnamefont {L.}~\bibnamefont {Tong}}, \bibinfo {author} {\bibfnamefont
  {H.}~\bibnamefont {Wang}}, \emph {et~al.},\ }\bibfield  {title} {\bibinfo
  {title} {{Ultrafast all-optical graphene modulator}},\ }\href
  {https://doi.org/10.1021/NL404356T/SUPPL_FILE/NL404356T_SI_001.PDF}
  {\bibfield  {journal} {\bibinfo  {journal} {Nano Letters}\ }\textbf {\bibinfo
  {volume} {14}},\ \bibinfo {pages} {955} (\bibinfo {year} {2014})}\BibitemShut
  {NoStop}%
\bibitem [{\citenamefont {Liu}\ \emph {et~al.}(2013)\citenamefont {Liu},
  \citenamefont {Feng}, \citenamefont {Jiang}, \citenamefont {Xin},
  \citenamefont {Wang}, \citenamefont {Sheng}, \citenamefont {Liu},
  \citenamefont {Wang}, \citenamefont {Zhou},\ and\ \citenamefont
  {Tian}}]{Liu2013}%
  \BibitemOpen
  \bibfield  {author} {\bibinfo {author} {\bibfnamefont {Z.~B.}\ \bibnamefont
  {Liu}}, \bibinfo {author} {\bibfnamefont {M.}~\bibnamefont {Feng}}, \bibinfo
  {author} {\bibfnamefont {W.~S.}\ \bibnamefont {Jiang}}, \bibinfo {author}
  {\bibfnamefont {W.}~\bibnamefont {Xin}}, \bibinfo {author} {\bibfnamefont
  {P.}~\bibnamefont {Wang}}, \bibinfo {author} {\bibfnamefont {Q.~W.}\
  \bibnamefont {Sheng}}, \bibinfo {author} {\bibfnamefont {Y.~G.}\ \bibnamefont
  {Liu}}, \bibinfo {author} {\bibfnamefont {D.~N.}\ \bibnamefont {Wang}},
  \bibinfo {author} {\bibfnamefont {W.~Y.}\ \bibnamefont {Zhou}},\ and\
  \bibinfo {author} {\bibfnamefont {J.~G.}\ \bibnamefont {Tian}},\ }\bibfield
  {title} {\bibinfo {title} {{Broadband all-optical modulation using a
  graphene-covered-microfiber}},\ }\href
  {https://doi.org/10.1088/1612-2011/10/6/065901} {\bibfield  {journal}
  {\bibinfo  {journal} {Laser Physics Letters}\ }\textbf {\bibinfo {volume}
  {10}},\ \bibinfo {pages} {065901} (\bibinfo {year} {2013})}\BibitemShut
  {NoStop}%
\bibitem [{\citenamefont {Heidari}\ \emph {et~al.}(2021)\citenamefont
  {Heidari}, \citenamefont {Faramarzi}, \citenamefont {Sharifi}, \citenamefont
  {Hashemi}, \citenamefont {Bahadori-Haghighi}, \citenamefont {Janjan},\ and\
  \citenamefont {Abbott}}]{Heidari2021}%
  \BibitemOpen
  \bibfield  {author} {\bibinfo {author} {\bibfnamefont {M.}~\bibnamefont
  {Heidari}}, \bibinfo {author} {\bibfnamefont {V.}~\bibnamefont {Faramarzi}},
  \bibinfo {author} {\bibfnamefont {Z.}~\bibnamefont {Sharifi}}, \bibinfo
  {author} {\bibfnamefont {M.}~\bibnamefont {Hashemi}}, \bibinfo {author}
  {\bibfnamefont {S.}~\bibnamefont {Bahadori-Haghighi}}, \bibinfo {author}
  {\bibfnamefont {B.}~\bibnamefont {Janjan}},\ and\ \bibinfo {author}
  {\bibfnamefont {D.}~\bibnamefont {Abbott}},\ }\bibfield  {title} {\bibinfo
  {title} {{A high-performance TE modulator/TM-pass polarizer using selective
  mode shaping in a VO2-based side-polished fiber}},\ }\href
  {https://doi.org/10.1515/NANOPH-2021-0225/ASSET/GRAPHIC/J_NANOPH-2021-0225_FIG_007.JPG}
  {\bibfield  {journal} {\bibinfo  {journal} {Nanophotonics}\ }\textbf
  {\bibinfo {volume} {10}},\ \bibinfo {pages} {3451} (\bibinfo {year}
  {2021})}\BibitemShut {NoStop}%
\bibitem [{\citenamefont {{Alexander Schmidt}}\ \emph
  {et~al.}(2016)\citenamefont {{Alexander Schmidt}}, \citenamefont {Argyros},
  \citenamefont {{Sorin A Schmidt}}, \citenamefont {{Schmidt Otto Schott}},
  \citenamefont {Schmidt}, \citenamefont {Argyros},\ and\ \citenamefont
  {Sorin}}]{AlexanderSchmidt2016}%
  \BibitemOpen
  \bibfield  {author} {\bibinfo {author} {\bibfnamefont {M.}~\bibnamefont
  {{Alexander Schmidt}}}, \bibinfo {author} {\bibfnamefont {A.}~\bibnamefont
  {Argyros}}, \bibinfo {author} {\bibfnamefont {F.~M.}\ \bibnamefont {{Sorin A
  Schmidt}}}, \bibinfo {author} {\bibfnamefont {M.~A.}\ \bibnamefont {{Schmidt
  Otto Schott}}}, \bibinfo {author} {\bibfnamefont {M.~A.}\ \bibnamefont
  {Schmidt}}, \bibinfo {author} {\bibfnamefont {A.}~\bibnamefont {Argyros}},\
  and\ \bibinfo {author} {\bibfnamefont {F.}~\bibnamefont {Sorin}},\ }\bibfield
   {title} {\bibinfo {title} {{Hybrid Optical Fibers – An Innovative Platform
  for In-Fiber Photonic Devices}},\ }\href
  {https://doi.org/10.1002/ADOM.201500319} {\bibfield  {journal} {\bibinfo
  {journal} {Advanced Optical Materials}\ }\textbf {\bibinfo {volume} {4}},\
  \bibinfo {pages} {13} (\bibinfo {year} {2016})}\BibitemShut {NoStop}%
\bibitem [{\citenamefont {Yu}\ \emph {et~al.}(2013)\citenamefont {Yu},
  \citenamefont {Zhang}, \citenamefont {Wang}, \citenamefont {Zhao},
  \citenamefont {Wang}, \citenamefont {Wen}, \citenamefont {Zhang},\ and\
  \citenamefont {Wang}}]{Yu2013}%
  \BibitemOpen
  \bibfield  {author} {\bibinfo {author} {\bibfnamefont {H.}~\bibnamefont
  {Yu}}, \bibinfo {author} {\bibfnamefont {H.}~\bibnamefont {Zhang}}, \bibinfo
  {author} {\bibfnamefont {Y.}~\bibnamefont {Wang}}, \bibinfo {author}
  {\bibfnamefont {C.}~\bibnamefont {Zhao}}, \bibinfo {author} {\bibfnamefont
  {B.}~\bibnamefont {Wang}}, \bibinfo {author} {\bibfnamefont {S.}~\bibnamefont
  {Wen}}, \bibinfo {author} {\bibfnamefont {H.}~\bibnamefont {Zhang}},\ and\
  \bibinfo {author} {\bibfnamefont {J.}~\bibnamefont {Wang}},\ }\bibfield
  {title} {\bibinfo {title} {{Topological insulator as an optical modulator for
  pulsed solid-state lasers}},\ }\href {https://doi.org/10.1002/LPOR.201300084}
  {\bibfield  {journal} {\bibinfo  {journal} {Laser \& Photonics Reviews}\
  }\textbf {\bibinfo {volume} {7}},\ \bibinfo {pages} {L77} (\bibinfo {year}
  {2013})}\BibitemShut {NoStop}%
\bibitem [{\citenamefont {Kou}\ \emph {et~al.}(2014)\citenamefont {Kou},
  \citenamefont {Chen}, \citenamefont {Chen}, \citenamefont {Xu},\ and\
  \citenamefont {Lu}}]{Kou2014}%
  \BibitemOpen
  \bibfield  {author} {\bibinfo {author} {\bibfnamefont {J.-L.}\ \bibnamefont
  {Kou}}, \bibinfo {author} {\bibfnamefont {J.-H.}\ \bibnamefont {Chen}},
  \bibinfo {author} {\bibfnamefont {Y.~E.}\ \bibnamefont {Chen}}, \bibinfo
  {author} {\bibfnamefont {F.}~\bibnamefont {Xu}},\ and\ \bibinfo {author}
  {\bibfnamefont {Y.-Q.}\ \bibnamefont {Lu}},\ }\bibfield  {title} {\bibinfo
  {title} {{Platform for enhanced light–graphene interaction length and
  miniaturizing fiber stereo devices}},\ }\href
  {https://doi.org/10.1364/OPTICA.1.000307} {\bibfield  {journal} {\bibinfo
  {journal} {Optica}\ }\textbf {\bibinfo {volume} {1}},\ \bibinfo {pages} {307}
  (\bibinfo {year} {2014})}\BibitemShut {NoStop}%
\bibitem [{\citenamefont {Bao}\ \emph {et~al.}(2011)\citenamefont {Bao},
  \citenamefont {Zhang}, \citenamefont {Wang}, \citenamefont {Ni},
  \citenamefont {Lim}, \citenamefont {Wang}, \citenamefont {Tang},\ and\
  \citenamefont {Loh}}]{Bao2011}%
  \BibitemOpen
  \bibfield  {author} {\bibinfo {author} {\bibfnamefont {Q.}~\bibnamefont
  {Bao}}, \bibinfo {author} {\bibfnamefont {H.}~\bibnamefont {Zhang}}, \bibinfo
  {author} {\bibfnamefont {B.}~\bibnamefont {Wang}}, \bibinfo {author}
  {\bibfnamefont {Z.}~\bibnamefont {Ni}}, \bibinfo {author} {\bibfnamefont
  {C.~H. Y.~X.}\ \bibnamefont {Lim}}, \bibinfo {author} {\bibfnamefont
  {Y.}~\bibnamefont {Wang}}, \bibinfo {author} {\bibfnamefont {D.~Y.}\
  \bibnamefont {Tang}},\ and\ \bibinfo {author} {\bibfnamefont {K.~P.}\
  \bibnamefont {Loh}},\ }\bibfield  {title} {\bibinfo {title} {{Broadband
  graphene polarizer}},\ }\href {https://doi.org/10.1038/nphoton.2011.102}
  {\bibfield  {journal} {\bibinfo  {journal} {Nature Photonics}\ }\textbf
  {\bibinfo {volume} {5}},\ \bibinfo {pages} {411} (\bibinfo {year}
  {2011})}\BibitemShut {NoStop}%
\bibitem [{\citenamefont {Li}\ \emph {et~al.}(2021)\citenamefont {Li},
  \citenamefont {Zhu}, \citenamefont {Zhan}, \citenamefont {Zhuo},
  \citenamefont {Che}, \citenamefont {Zhang}, \citenamefont {Zheng},
  \citenamefont {Tang}, \citenamefont {Zhang}, \citenamefont {Yu} \emph
  {et~al.}}]{Li2021}%
  \BibitemOpen
  \bibfield  {author} {\bibinfo {author} {\bibfnamefont {D.}~\bibnamefont
  {Li}}, \bibinfo {author} {\bibfnamefont {W.}~\bibnamefont {Zhu}}, \bibinfo
  {author} {\bibfnamefont {Y.}~\bibnamefont {Zhan}}, \bibinfo {author}
  {\bibfnamefont {L.}~\bibnamefont {Zhuo}}, \bibinfo {author} {\bibfnamefont
  {Z.}~\bibnamefont {Che}}, \bibinfo {author} {\bibfnamefont {Y.}~\bibnamefont
  {Zhang}}, \bibinfo {author} {\bibfnamefont {H.}~\bibnamefont {Zheng}},
  \bibinfo {author} {\bibfnamefont {J.}~\bibnamefont {Tang}}, \bibinfo {author}
  {\bibfnamefont {J.}~\bibnamefont {Zhang}}, \bibinfo {author} {\bibfnamefont
  {J.}~\bibnamefont {Yu}}, \emph {et~al.},\ }\bibfield  {title} {\bibinfo
  {title} {{Local-field-enhanced and polarisation-sensitive graphene/MoS2 film
  on side-polished fibre with coated Au film}},\ }\href
  {https://doi.org/10.1016/j.optcom.2021.126966} {\bibfield  {journal}
  {\bibinfo  {journal} {Optics Communications}\ }\textbf {\bibinfo {volume}
  {491}},\ \bibinfo {pages} {126966} (\bibinfo {year} {2021})}\BibitemShut
  {NoStop}%
\bibitem [{\citenamefont {Chu}\ \emph {et~al.}(2017)\citenamefont {Chu},
  \citenamefont {Guan}, \citenamefont {Yang}, \citenamefont {Zhu},
  \citenamefont {Shi}, \citenamefont {Tian}, \citenamefont {Yuan},
  \citenamefont {Brambilla}, \citenamefont {Yin}, \citenamefont {Ke} \emph
  {et~al.}}]{Chu2017}%
  \BibitemOpen
  \bibfield  {author} {\bibinfo {author} {\bibfnamefont {R.}~\bibnamefont
  {Chu}}, \bibinfo {author} {\bibfnamefont {C.}~\bibnamefont {Guan}}, \bibinfo
  {author} {\bibfnamefont {J.}~\bibnamefont {Yang}}, \bibinfo {author}
  {\bibfnamefont {Z.}~\bibnamefont {Zhu}}, \bibinfo {author} {\bibfnamefont
  {J.}~\bibnamefont {Shi}}, \bibinfo {author} {\bibfnamefont {P.}~\bibnamefont
  {Tian}}, \bibinfo {author} {\bibfnamefont {L.}~\bibnamefont {Yuan}}, \bibinfo
  {author} {\bibfnamefont {G.}~\bibnamefont {Brambilla}}, \bibinfo {author}
  {\bibfnamefont {X.}~\bibnamefont {Yin}}, \bibinfo {author} {\bibfnamefont
  {X.}~\bibnamefont {Ke}}, \emph {et~al.},\ }\bibfield  {title} {\bibinfo
  {title} {{High extinction ratio D-shaped fiber polarizers coated by a double
  graphene/PMMA stack}},\ }\href {https://doi.org/10.1364/OE.25.013278}
  {\bibfield  {journal} {\bibinfo  {journal} {Optics Express}\ }\textbf
  {\bibinfo {volume} {25}},\ \bibinfo {pages} {13278} (\bibinfo {year}
  {2017})}\BibitemShut {NoStop}%
\bibitem [{\citenamefont {Nikbakht}\ \emph {et~al.}(2019)\citenamefont
  {Nikbakht}, \citenamefont {Latifi}, \citenamefont {Parsanasab}, \citenamefont
  {Taghavi},\ and\ \citenamefont {Riyahi}}]{Nikbakht2019}%
  \BibitemOpen
  \bibfield  {author} {\bibinfo {author} {\bibfnamefont {H.}~\bibnamefont
  {Nikbakht}}, \bibinfo {author} {\bibfnamefont {H.}~\bibnamefont {Latifi}},
  \bibinfo {author} {\bibfnamefont {G.~M.}\ \bibnamefont {Parsanasab}},
  \bibinfo {author} {\bibfnamefont {M.}~\bibnamefont {Taghavi}},\ and\ \bibinfo
  {author} {\bibfnamefont {M.}~\bibnamefont {Riyahi}},\ }\bibfield  {title}
  {\bibinfo {title} {{Utilizing polarization-selective mode shaping by
  chalcogenide thin film to enhance the performance of graphene-based
  integrated optical devices}},\ }\href
  {https://doi.org/10.1038/s41598-019-48890-y} {\bibfield  {journal} {\bibinfo
  {journal} {Scientific Reports}\ }\textbf {\bibinfo {volume} {9}},\ \bibinfo
  {pages} {1} (\bibinfo {year} {2019})}\BibitemShut {NoStop}%
\bibitem [{\citenamefont {Li}\ \emph {et~al.}(2015)\citenamefont {Li},
  \citenamefont {Wang}, \citenamefont {Sun}, \citenamefont {Duan},
  \citenamefont {Wang},\ and\ \citenamefont {Si}}]{Li2015}%
  \BibitemOpen
  \bibfield  {author} {\bibinfo {author} {\bibfnamefont {L.}~\bibnamefont
  {Li}}, \bibinfo {author} {\bibfnamefont {Y.}~\bibnamefont {Wang}}, \bibinfo
  {author} {\bibfnamefont {H.}~\bibnamefont {Sun}}, \bibinfo {author}
  {\bibfnamefont {L.}~\bibnamefont {Duan}}, \bibinfo {author} {\bibfnamefont
  {X.}~\bibnamefont {Wang}},\ and\ \bibinfo {author} {\bibfnamefont
  {J.}~\bibnamefont {Si}},\ }\bibfield  {title} {\bibinfo {title}
  {{Single-walled carbon nanotube solution-based saturable absorbers for
  mode-locked fiber laser}},\ }\href {https://doi.org/10.1117/1.oe.54.8.086103}
  {\bibfield  {journal} {\bibinfo  {journal} {Optical Engineering}\ }\textbf
  {\bibinfo {volume} {54}},\ \bibinfo {pages} {086103} (\bibinfo {year}
  {2015})}\BibitemShut {NoStop}%
\bibitem [{\citenamefont {Song}\ \emph {et~al.}(2007)\citenamefont {Song},
  \citenamefont {Yamashita}, \citenamefont {Goh},\ and\ \citenamefont
  {Set}}]{Song2007}%
  \BibitemOpen
  \bibfield  {author} {\bibinfo {author} {\bibfnamefont {Y.-W.}\ \bibnamefont
  {Song}}, \bibinfo {author} {\bibfnamefont {S.}~\bibnamefont {Yamashita}},
  \bibinfo {author} {\bibfnamefont {C.~S.}\ \bibnamefont {Goh}},\ and\ \bibinfo
  {author} {\bibfnamefont {S.~Y.}\ \bibnamefont {Set}},\ }\bibfield  {title}
  {\bibinfo {title} {{Carbon nanotube mode lockers with enhanced nonlinearity
  via evanescent field interaction in D-shaped fibers}},\ }\href
  {https://doi.org/10.1364/ol.32.000148} {\bibfield  {journal} {\bibinfo
  {journal} {Optics Letters}\ }\textbf {\bibinfo {volume} {32}},\ \bibinfo
  {pages} {148} (\bibinfo {year} {2007})}\BibitemShut {NoStop}%
\bibitem [{\citenamefont {Jeong}\ \emph {et~al.}(2016)\citenamefont {Jeong},
  \citenamefont {Choi}, \citenamefont {Rotermund}, \citenamefont {Lee},\ and\
  \citenamefont {Yeom}}]{Jeong2016}%
  \BibitemOpen
  \bibfield  {author} {\bibinfo {author} {\bibfnamefont {H.}~\bibnamefont
  {Jeong}}, \bibinfo {author} {\bibfnamefont {S.~Y.}\ \bibnamefont {Choi}},
  \bibinfo {author} {\bibfnamefont {F.}~\bibnamefont {Rotermund}}, \bibinfo
  {author} {\bibfnamefont {K.}~\bibnamefont {Lee}},\ and\ \bibinfo {author}
  {\bibfnamefont {D.~I.}\ \bibnamefont {Yeom}},\ }\bibfield  {title} {\bibinfo
  {title} {{All-Polarization Maintaining Passively Mode-Locked Fiber Laser
  Using Evanescent Field Interaction With Single-Walled Carbon Nanotube
  Saturable Absorber}},\ }\href {https://doi.org/10.1109/JLT.2016.2543754}
  {\bibfield  {journal} {\bibinfo  {journal} {Journal of Lightwave Technology}\
  }\textbf {\bibinfo {volume} {34}},\ \bibinfo {pages} {3510} (\bibinfo {year}
  {2016})}\BibitemShut {NoStop}%
\bibitem [{\citenamefont {Lee}\ \emph {et~al.}(2017)\citenamefont {Lee},
  \citenamefont {Koo}, \citenamefont {Lee}, \citenamefont {{Min Jhon}},
  \citenamefont {{Han Lee}}, \citenamefont {McAleavey}, \citenamefont
  {Donegan}, \citenamefont {MacCraith}, \citenamefont {Hegarty}, \citenamefont
  {Maz{\'{e}}} \emph {et~al.}}]{Lee2017}%
  \BibitemOpen
  \bibfield  {author} {\bibinfo {author} {\bibfnamefont {J.}~\bibnamefont
  {Lee}}, \bibinfo {author} {\bibfnamefont {J.}~\bibnamefont {Koo}}, \bibinfo
  {author} {\bibfnamefont {J.}~\bibnamefont {Lee}}, \bibinfo {author}
  {\bibfnamefont {Y.}~\bibnamefont {{Min Jhon}}}, \bibinfo {author}
  {\bibfnamefont {J.}~\bibnamefont {{Han Lee}}}, \bibinfo {author}
  {\bibfnamefont {F.~J.}\ \bibnamefont {McAleavey}}, \bibinfo {author}
  {\bibfnamefont {J.~F.}\ \bibnamefont {Donegan}}, \bibinfo {author}
  {\bibfnamefont {B.~D.}\ \bibnamefont {MacCraith}}, \bibinfo {author}
  {\bibfnamefont {J.}~\bibnamefont {Hegarty}}, \bibinfo {author} {\bibfnamefont
  {G.}~\bibnamefont {Maz{\'{e}}}}, \emph {et~al.},\ }\bibfield  {title}
  {\bibinfo {title} {{All-fiberized, femtosecond laser at 1912 nm using a
  bulk-like MoSe2 saturable absorber}},\ }\href
  {https://doi.org/10.1364/OME.7.002968} {\bibfield  {journal} {\bibinfo
  {journal} {Optical Materials Express}\ }\textbf {\bibinfo {volume} {7}},\
  \bibinfo {pages} {2968} (\bibinfo {year} {2017})}\BibitemShut {NoStop}%
\bibitem [{\citenamefont {Lee}\ \emph {et~al.}(2015)\citenamefont {Lee},
  \citenamefont {Koo}, \citenamefont {{Min Jhon}}, \citenamefont {{Han Lee}},
  \citenamefont {R{\"{o}}ser}, \citenamefont {Rothhard}, \citenamefont {Ortac},
  \citenamefont {Liem}, \citenamefont {Schmidt}, \citenamefont {Schreiber}
  \emph {et~al.}}]{Lee2015}%
  \BibitemOpen
  \bibfield  {author} {\bibinfo {author} {\bibfnamefont {J.}~\bibnamefont
  {Lee}}, \bibinfo {author} {\bibfnamefont {J.}~\bibnamefont {Koo}}, \bibinfo
  {author} {\bibfnamefont {Y.}~\bibnamefont {{Min Jhon}}}, \bibinfo {author}
  {\bibfnamefont {J.}~\bibnamefont {{Han Lee}}}, \bibinfo {author}
  {\bibfnamefont {F.}~\bibnamefont {R{\"{o}}ser}}, \bibinfo {author}
  {\bibfnamefont {J.}~\bibnamefont {Rothhard}}, \bibinfo {author}
  {\bibfnamefont {B.}~\bibnamefont {Ortac}}, \bibinfo {author} {\bibfnamefont
  {A.}~\bibnamefont {Liem}}, \bibinfo {author} {\bibfnamefont {O.}~\bibnamefont
  {Schmidt}}, \bibinfo {author} {\bibfnamefont {T.}~\bibnamefont {Schreiber}},
  \emph {et~al.},\ }\bibfield  {title} {\bibinfo {title} {{Femtosecond harmonic
  mode-locking of a fiber laser based on a bulk-structured Bi2Te3 topological
  insulator}},\ }\href {https://doi.org/10.1364/OE.23.006359} {\bibfield
  {journal} {\bibinfo  {journal} {Optics Express}\ }\textbf {\bibinfo {volume}
  {23}},\ \bibinfo {pages} {6359} (\bibinfo {year} {2015})}\BibitemShut
  {NoStop}%
\bibitem [{\citenamefont {Zhao}\ \emph {et~al.}(2012)\citenamefont {Zhao},
  \citenamefont {Zhang}, \citenamefont {Qi}, \citenamefont {Chen},
  \citenamefont {Wang}, \citenamefont {Wen},\ and\ \citenamefont
  {Tang}}]{Zhao2012}%
  \BibitemOpen
  \bibfield  {author} {\bibinfo {author} {\bibfnamefont {C.}~\bibnamefont
  {Zhao}}, \bibinfo {author} {\bibfnamefont {H.}~\bibnamefont {Zhang}},
  \bibinfo {author} {\bibfnamefont {X.}~\bibnamefont {Qi}}, \bibinfo {author}
  {\bibfnamefont {Y.}~\bibnamefont {Chen}}, \bibinfo {author} {\bibfnamefont
  {Z.}~\bibnamefont {Wang}}, \bibinfo {author} {\bibfnamefont {S.}~\bibnamefont
  {Wen}},\ and\ \bibinfo {author} {\bibfnamefont {D.}~\bibnamefont {Tang}},\
  }\bibfield  {title} {\bibinfo {title} {{Ultra-short pulse generation by a
  topological insulator based saturable absorber}},\ }\href
  {https://doi.org/10.1063/1.4767919} {\bibfield  {journal} {\bibinfo
  {journal} {Applied Physics Letters}\ }\textbf {\bibinfo {volume} {101}},\
  \bibinfo {pages} {211106} (\bibinfo {year} {2012})}\BibitemShut {NoStop}%
\bibitem [{\citenamefont {Zapata}\ \emph {et~al.}(2016)\citenamefont {Zapata},
  \citenamefont {Steinberg}, \citenamefont {Saito}, \citenamefont {{De
  Oliveira}}, \citenamefont {C{\'{a}}rdenas},\ and\ \citenamefont {{De
  Souza}}}]{Zapata2016}%
  \BibitemOpen
  \bibfield  {author} {\bibinfo {author} {\bibfnamefont {J.~D.}\ \bibnamefont
  {Zapata}}, \bibinfo {author} {\bibfnamefont {D.}~\bibnamefont {Steinberg}},
  \bibinfo {author} {\bibfnamefont {L.~A.}\ \bibnamefont {Saito}}, \bibinfo
  {author} {\bibfnamefont {R.~E.}\ \bibnamefont {{De Oliveira}}}, \bibinfo
  {author} {\bibfnamefont {A.~M.}\ \bibnamefont {C{\'{a}}rdenas}},\ and\
  \bibinfo {author} {\bibfnamefont {E.~A.}\ \bibnamefont {{De Souza}}},\
  }\bibfield  {title} {\bibinfo {title} {{Efficient graphene saturable
  absorbers on D-shaped optical fiber for ultrashort pulse generation}},\
  }\href {https://doi.org/10.1038/srep20644} {\bibfield  {journal} {\bibinfo
  {journal} {Scientific Reports}\ }\textbf {\bibinfo {volume} {6}},\ \bibinfo
  {pages} {1} (\bibinfo {year} {2016})}\BibitemShut {NoStop}%
\bibitem [{\citenamefont {Jiang}\ \emph {et~al.}(2022)\citenamefont {Jiang},
  \citenamefont {Zhou}, \citenamefont {Lou}, \citenamefont {Liu}, \citenamefont
  {Xu}, \citenamefont {Zhao}, \citenamefont {Li}, \citenamefont {Tang},\ and\
  \citenamefont {Shen}}]{Jiang2022}%
  \BibitemOpen
  \bibfield  {author} {\bibinfo {author} {\bibfnamefont {Y.}~\bibnamefont
  {Jiang}}, \bibinfo {author} {\bibfnamefont {J.}~\bibnamefont {Zhou}},
  \bibinfo {author} {\bibfnamefont {B.}~\bibnamefont {Lou}}, \bibinfo {author}
  {\bibfnamefont {J.}~\bibnamefont {Liu}}, \bibinfo {author} {\bibfnamefont
  {Y.}~\bibnamefont {Xu}}, \bibinfo {author} {\bibfnamefont {J.}~\bibnamefont
  {Zhao}}, \bibinfo {author} {\bibfnamefont {L.}~\bibnamefont {Li}}, \bibinfo
  {author} {\bibfnamefont {D.}~\bibnamefont {Tang}},\ and\ \bibinfo {author}
  {\bibfnamefont {D.}~\bibnamefont {Shen}},\ }\bibfield  {title} {\bibinfo
  {title} {{Local nonlinearity engineering of evanescent-field-interaction
  fiber devices embedding in black phosphorus quantum dots}},\ }\href
  {https://doi.org/10.1515/NANOPH-2021-0513/ASSET/GRAPHIC/J_NANOPH-2021-0513_FIG_010.JPG}
  {\bibfield  {journal} {\bibinfo  {journal} {Nanophotonics}\ }\textbf
  {\bibinfo {volume} {11}},\ \bibinfo {pages} {87} (\bibinfo {year}
  {2022})}\BibitemShut {NoStop}%
\bibitem [{\citenamefont {Chen}\ \emph {et~al.}(2021)\citenamefont {Chen},
  \citenamefont {Xiong}, \citenamefont {Xu},\ and\ \citenamefont
  {Lu}}]{Chen2021}%
  \BibitemOpen
  \bibfield  {author} {\bibinfo {author} {\bibfnamefont {J.-H.}\ \bibnamefont
  {Chen}}, \bibinfo {author} {\bibfnamefont {Y.-F.}\ \bibnamefont {Xiong}},
  \bibinfo {author} {\bibfnamefont {F.}~\bibnamefont {Xu}},\ and\ \bibinfo
  {author} {\bibfnamefont {Y.-Q.}\ \bibnamefont {Lu}},\ }\bibfield  {title}
  {\bibinfo {title} {{Silica optical fiber integrated with two-dimensional
  materials: towards opto-electro-mechanical technology}},\ }\href
  {https://doi.org/10.1038/s41377-021-00520-x} {\bibfield  {journal} {\bibinfo
  {journal} {Light: Science \& Applications}\ }\textbf {\bibinfo {volume}
  {10}},\ \bibinfo {pages} {78} (\bibinfo {year} {2021})}\BibitemShut {NoStop}%
\bibitem [{Dem(2019)}]{Demongodin2019}%
  \BibitemOpen
  \bibfield  {title} {\bibinfo {title} {{Ultrafast saturable absorption
  dynamics in hybrid graphene/Si3N4 waveguides}},\ }\href
  {https://doi.org/10.1063/1.5094523} {\bibfield  {journal} {\bibinfo
  {journal} {APL Photonics}\ }\textbf {\bibinfo {volume} {4}},\ \bibinfo
  {pages} {076102} (\bibinfo {year} {2019})}\BibitemShut {NoStop}%
\bibitem [{\citenamefont {Sederberg}\ and\ \citenamefont
  {Elezzabi}(2014)}]{Sederberg2014}%
  \BibitemOpen
  \bibfield  {author} {\bibinfo {author} {\bibfnamefont {S.}~\bibnamefont
  {Sederberg}}\ and\ \bibinfo {author} {\bibfnamefont {A.~Y.}\ \bibnamefont
  {Elezzabi}},\ }\bibfield  {title} {\bibinfo {title} {{Nonmonotonic
  Wavelength-Dependent Power Scaling in Silicon-on-Insulator Waveguides via
  Nonlinear Optical Effect Conglomeration}},\ }\href
  {https://doi.org/10.1021/PH500022F/SUPPL_FILE/PH500022F_SI_001.PDF}
  {\bibfield  {journal} {\bibinfo  {journal} {ACS Photonics}\ }\textbf
  {\bibinfo {volume} {1}},\ \bibinfo {pages} {576} (\bibinfo {year}
  {2014})}\BibitemShut {NoStop}%
\bibitem [{\citenamefont {Park}\ \emph {et~al.}(2015)\citenamefont {Park},
  \citenamefont {Jeong}, \citenamefont {Choi}, \citenamefont {Kim},
  \citenamefont {Rotermund},\ and\ \citenamefont {Yeom}}]{Park2015}%
  \BibitemOpen
  \bibfield  {author} {\bibinfo {author} {\bibfnamefont {N.~H.}\ \bibnamefont
  {Park}}, \bibinfo {author} {\bibfnamefont {H.}~\bibnamefont {Jeong}},
  \bibinfo {author} {\bibfnamefont {S.~Y.}\ \bibnamefont {Choi}}, \bibinfo
  {author} {\bibfnamefont {M.~H.}\ \bibnamefont {Kim}}, \bibinfo {author}
  {\bibfnamefont {F.}~\bibnamefont {Rotermund}},\ and\ \bibinfo {author}
  {\bibfnamefont {D.-I.}\ \bibnamefont {Yeom}},\ }\bibfield  {title} {\bibinfo
  {title} {{Monolayer graphene saturable absorbers with strongly enhanced
  evanescent-field interaction for ultrafast fiber laser mode-locking}},\
  }\href {https://doi.org/10.1364/oe.23.019806} {\bibfield  {journal} {\bibinfo
   {journal} {Optics Express}\ }\textbf {\bibinfo {volume} {23}},\ \bibinfo
  {pages} {19806} (\bibinfo {year} {2015})}\BibitemShut {NoStop}%
\bibitem [{\citenamefont {Mkrtchyan}\ \emph {et~al.}(2019)\citenamefont
  {Mkrtchyan}, \citenamefont {Gladush}, \citenamefont {Galiakhmetova},
  \citenamefont {Yakovlev}, \citenamefont {Ahtyamov},\ and\ \citenamefont
  {Nasibulin}}]{Mkrtchyan2019}%
  \BibitemOpen
  \bibfield  {author} {\bibinfo {author} {\bibfnamefont {A.~A.}\ \bibnamefont
  {Mkrtchyan}}, \bibinfo {author} {\bibfnamefont {Y.~G.}\ \bibnamefont
  {Gladush}}, \bibinfo {author} {\bibfnamefont {D.~I.}\ \bibnamefont
  {Galiakhmetova}}, \bibinfo {author} {\bibfnamefont {V.~Y.}\ \bibnamefont
  {Yakovlev}}, \bibinfo {author} {\bibfnamefont {V.~T.}\ \bibnamefont
  {Ahtyamov}},\ and\ \bibinfo {author} {\bibfnamefont {A.~G.}\ \bibnamefont
  {Nasibulin}},\ }\bibfield  {title} {\bibinfo {title} {{Dry-transfer technique
  for polymer-free single-walled carbon nanotube saturable absorber on a side
  polished fiber}},\ }\href {https://doi.org/10.1364/OME.9.001551} {\bibfield
  {journal} {\bibinfo  {journal} {Optical Materials Express}\ }\textbf
  {\bibinfo {volume} {9}},\ \bibinfo {pages} {1551} (\bibinfo {year}
  {2019})}\BibitemShut {NoStop}%
\bibitem [{\citenamefont {Mustonen}\ \emph {et~al.}(2015)\citenamefont
  {Mustonen}, \citenamefont {Laiho}, \citenamefont {Kaskela}, \citenamefont
  {Susi}, \citenamefont {Nasibulin},\ and\ \citenamefont
  {Kauppinen}}]{Mustonen2015}%
  \BibitemOpen
  \bibfield  {author} {\bibinfo {author} {\bibfnamefont {K.}~\bibnamefont
  {Mustonen}}, \bibinfo {author} {\bibfnamefont {P.}~\bibnamefont {Laiho}},
  \bibinfo {author} {\bibfnamefont {A.}~\bibnamefont {Kaskela}}, \bibinfo
  {author} {\bibfnamefont {T.}~\bibnamefont {Susi}}, \bibinfo {author}
  {\bibfnamefont {A.~G.}\ \bibnamefont {Nasibulin}},\ and\ \bibinfo {author}
  {\bibfnamefont {E.~I.}\ \bibnamefont {Kauppinen}},\ }\bibfield  {title}
  {\bibinfo {title} {{Uncovering the ultimate performance of single-walled
  carbon nanotube films as transparent conductors}},\ }\href
  {https://doi.org/10.1063/1.4932942} {\bibfield  {journal} {\bibinfo
  {journal} {Applied Physics Letters}\ }\textbf {\bibinfo {volume} {107}},\
  \bibinfo {pages} {143113} (\bibinfo {year} {2015})}\BibitemShut {NoStop}%
\bibitem [{\citenamefont {Tian}\ \emph {et~al.}(2011)\citenamefont {Tian},
  \citenamefont {Nasibulin}, \citenamefont {Aitchison}, \citenamefont
  {Nikitin}, \citenamefont {Pfaler}, \citenamefont {Jiang}, \citenamefont
  {Zhu}, \citenamefont {Khriachtchev}, \citenamefont {Brown},\ and\
  \citenamefont {Kauppinen}}]{Tian2011}%
  \BibitemOpen
  \bibfield  {author} {\bibinfo {author} {\bibfnamefont {Y.}~\bibnamefont
  {Tian}}, \bibinfo {author} {\bibfnamefont {A.~G.}\ \bibnamefont {Nasibulin}},
  \bibinfo {author} {\bibfnamefont {B.}~\bibnamefont {Aitchison}}, \bibinfo
  {author} {\bibfnamefont {T.}~\bibnamefont {Nikitin}}, \bibinfo {author}
  {\bibfnamefont {J.~V.}\ \bibnamefont {Pfaler}}, \bibinfo {author}
  {\bibfnamefont {H.}~\bibnamefont {Jiang}}, \bibinfo {author} {\bibfnamefont
  {Z.}~\bibnamefont {Zhu}}, \bibinfo {author} {\bibfnamefont {L.}~\bibnamefont
  {Khriachtchev}}, \bibinfo {author} {\bibfnamefont {D.~P.}\ \bibnamefont
  {Brown}},\ and\ \bibinfo {author} {\bibfnamefont {E.~I.}\ \bibnamefont
  {Kauppinen}},\ }\bibfield  {title} {\bibinfo {title} {{Controlled synthesis
  of single-walled carbon nanotubes in an aerosol reactor}},\ }\href
  {https://doi.org/10.1021/jp112291f} {\bibfield  {journal} {\bibinfo
  {journal} {Journal of Physical Chemistry C}\ }\textbf {\bibinfo {volume}
  {115}},\ \bibinfo {pages} {7309} (\bibinfo {year} {2011})}\BibitemShut
  {NoStop}%
\bibitem [{\citenamefont {Nasibulin}\ \emph {et~al.}(2006)\citenamefont
  {Nasibulin}, \citenamefont {Brown}, \citenamefont {Queipo}, \citenamefont
  {Gonzalez}, \citenamefont {Jiang},\ and\ \citenamefont
  {Kauppinen}}]{Nasibulin2006}%
  \BibitemOpen
  \bibfield  {author} {\bibinfo {author} {\bibfnamefont {A.~G.}\ \bibnamefont
  {Nasibulin}}, \bibinfo {author} {\bibfnamefont {D.~P.}\ \bibnamefont
  {Brown}}, \bibinfo {author} {\bibfnamefont {P.}~\bibnamefont {Queipo}},
  \bibinfo {author} {\bibfnamefont {D.}~\bibnamefont {Gonzalez}}, \bibinfo
  {author} {\bibfnamefont {H.}~\bibnamefont {Jiang}},\ and\ \bibinfo {author}
  {\bibfnamefont {E.~I.}\ \bibnamefont {Kauppinen}},\ }\bibfield  {title}
  {\bibinfo {title} {{An essential role of CO2 and H2O during single-walled CNT
  synthesis from carbon monoxide}},\ }\href
  {https://doi.org/10.1016/j.cplett.2005.10.022} {\bibfield  {journal}
  {\bibinfo  {journal} {Chemical Physics Letters}\ }\textbf {\bibinfo {volume}
  {417}},\ \bibinfo {pages} {179} (\bibinfo {year} {2006})}\BibitemShut
  {NoStop}%
\bibitem [{\citenamefont {Ermolaev}\ \emph {et~al.}(2020)\citenamefont
  {Ermolaev}, \citenamefont {Tsapenko}, \citenamefont {Volkov}, \citenamefont
  {Anisimov}, \citenamefont {Gladush},\ and\ \citenamefont
  {Nasibulin}}]{Ermolaev2020}%
  \BibitemOpen
  \bibfield  {author} {\bibinfo {author} {\bibfnamefont {G.~A.}\ \bibnamefont
  {Ermolaev}}, \bibinfo {author} {\bibfnamefont {A.~P.}\ \bibnamefont
  {Tsapenko}}, \bibinfo {author} {\bibfnamefont {V.~S.}\ \bibnamefont
  {Volkov}}, \bibinfo {author} {\bibfnamefont {A.~S.}\ \bibnamefont
  {Anisimov}}, \bibinfo {author} {\bibfnamefont {Y.~G.}\ \bibnamefont
  {Gladush}},\ and\ \bibinfo {author} {\bibfnamefont {A.~G.}\ \bibnamefont
  {Nasibulin}},\ }\bibfield  {title} {\bibinfo {title} {{Express determination
  of thickness and dielectric function of single-walled carbon nanotube
  films}},\ }\href {https://doi.org/10.1063/5.0012933} {\bibfield  {journal}
  {\bibinfo  {journal} {Applied Physics Letters}\ }\textbf {\bibinfo {volume}
  {116}},\ \bibinfo {pages} {231103} (\bibinfo {year} {2020})}\BibitemShut
  {NoStop}%
\bibitem [{\citenamefont {Gladush}\ \emph {et~al.}(2019)\citenamefont
  {Gladush}, \citenamefont {Mkrtchyan}, \citenamefont {Kopylova}, \citenamefont
  {Ivanenko}, \citenamefont {Nyushkov}, \citenamefont {Kobtsev}, \citenamefont
  {Kokhanovskiy}, \citenamefont {Khegai}, \citenamefont {Melkumov},
  \citenamefont {Burdanova} \emph {et~al.}}]{Gladush2019}%
  \BibitemOpen
  \bibfield  {author} {\bibinfo {author} {\bibfnamefont {Y.}~\bibnamefont
  {Gladush}}, \bibinfo {author} {\bibfnamefont {A.~A.}\ \bibnamefont
  {Mkrtchyan}}, \bibinfo {author} {\bibfnamefont {D.~S.}\ \bibnamefont
  {Kopylova}}, \bibinfo {author} {\bibfnamefont {A.}~\bibnamefont {Ivanenko}},
  \bibinfo {author} {\bibfnamefont {B.}~\bibnamefont {Nyushkov}}, \bibinfo
  {author} {\bibfnamefont {S.}~\bibnamefont {Kobtsev}}, \bibinfo {author}
  {\bibfnamefont {A.}~\bibnamefont {Kokhanovskiy}}, \bibinfo {author}
  {\bibfnamefont {A.}~\bibnamefont {Khegai}}, \bibinfo {author} {\bibfnamefont
  {M.}~\bibnamefont {Melkumov}}, \bibinfo {author} {\bibfnamefont
  {M.}~\bibnamefont {Burdanova}}, \emph {et~al.},\ }\bibfield  {title}
  {\bibinfo {title} {{Ionic Liquid Gated Carbon Nanotube Saturable Absorber for
  Switchable Pulse Generation}},\ }\href
  {https://doi.org/10.1021/acs.nanolett.9b01012} {\bibfield  {journal}
  {\bibinfo  {journal} {Nano Letters}\ }\textbf {\bibinfo {volume} {19}},\
  \bibinfo {pages} {5836} (\bibinfo {year} {2019})}\BibitemShut {NoStop}%
\bibitem [{\citenamefont {Kataura}\ \emph {et~al.}(1999)\citenamefont
  {Kataura}, \citenamefont {Achiba},\ and\ \citenamefont
  {Jacquemin}}]{Kataura1999}%
  \BibitemOpen
  \bibfield  {author} {\bibinfo {author} {\bibfnamefont {H.}~\bibnamefont
  {Kataura}}, \bibinfo {author} {\bibfnamefont {Y.}~\bibnamefont {Achiba}},\
  and\ \bibinfo {author} {\bibfnamefont {R.}~\bibnamefont {Jacquemin}},\
  }\bibfield  {title} {\bibinfo {title} {{Amphoteric doping of single-wall
  carbon-nanotube thin films as probed by optical absorption spectroscopy}},\
  }\href {https://doi.org/10.1103/PhysRevB.60.13339} {\bibfield  {journal}
  {\bibinfo  {journal} {Physical Review B}\ }\textbf {\bibinfo {volume} {60}},\
  \bibinfo {pages} {13339} (\bibinfo {year} {1999})}\BibitemShut {NoStop}%
\bibitem [{\citenamefont {Ichida}\ \emph {et~al.}(2004)\citenamefont {Ichida},
  \citenamefont {Mizuno}, \citenamefont {Kataura}, \citenamefont {Achiba},\
  and\ \citenamefont {Nakamura}}]{Ichida2004}%
  \BibitemOpen
  \bibfield  {author} {\bibinfo {author} {\bibfnamefont {M.}~\bibnamefont
  {Ichida}}, \bibinfo {author} {\bibfnamefont {S.}~\bibnamefont {Mizuno}},
  \bibinfo {author} {\bibfnamefont {H.}~\bibnamefont {Kataura}}, \bibinfo
  {author} {\bibfnamefont {Y.}~\bibnamefont {Achiba}},\ and\ \bibinfo {author}
  {\bibfnamefont {A.}~\bibnamefont {Nakamura}},\ }\bibfield  {title} {\bibinfo
  {title} {{Anisotropic optical properties of mechanically aligned
  single-walled carbon nanotubes in polymer}},\ }\href
  {https://doi.org/10.1007/S00339-003-2462-4} {\bibfield  {journal} {\bibinfo
  {journal} {Applied Physics A}\ }\textbf {\bibinfo {volume} {78}},\ \bibinfo
  {pages} {1117} (\bibinfo {year} {2004})}\BibitemShut {NoStop}%
\bibitem [{\citenamefont {Maeda}\ \emph {et~al.}(2005)\citenamefont {Maeda},
  \citenamefont {Matsumoto}, \citenamefont {Kishida}, \citenamefont {Takenobu},
  \citenamefont {Iwasa}, \citenamefont {Shiraishi}, \citenamefont {Ata},\ and\
  \citenamefont {Okamoto}}]{Maeda2005}%
  \BibitemOpen
  \bibfield  {author} {\bibinfo {author} {\bibfnamefont {A.}~\bibnamefont
  {Maeda}}, \bibinfo {author} {\bibfnamefont {S.}~\bibnamefont {Matsumoto}},
  \bibinfo {author} {\bibfnamefont {H.}~\bibnamefont {Kishida}}, \bibinfo
  {author} {\bibfnamefont {T.}~\bibnamefont {Takenobu}}, \bibinfo {author}
  {\bibfnamefont {Y.}~\bibnamefont {Iwasa}}, \bibinfo {author} {\bibfnamefont
  {M.}~\bibnamefont {Shiraishi}}, \bibinfo {author} {\bibfnamefont
  {M.}~\bibnamefont {Ata}},\ and\ \bibinfo {author} {\bibfnamefont
  {H.}~\bibnamefont {Okamoto}},\ }\bibfield  {title} {\bibinfo {title} {{Large
  Optical Nonlinearity of Semiconducting Single-Walled Carbon Nanotubes under
  Resonant Excitations}},\ }\href
  {https://doi.org/10.1103/PhysRevLett.94.047404} {\bibfield  {journal}
  {\bibinfo  {journal} {Physical Review Letters}\ }\textbf {\bibinfo {volume}
  {94}},\ \bibinfo {pages} {047404} (\bibinfo {year} {2005})}\BibitemShut
  {NoStop}%
\bibitem [{\citenamefont {Xu}\ \emph {et~al.}(2016)\citenamefont {Xu},
  \citenamefont {Wang}, \citenamefont {Zhu}, \citenamefont {Meng},
  \citenamefont {Liu}, \citenamefont {Liu}, \citenamefont {Tang}, \citenamefont
  {Liu}, \citenamefont {Hu}, \citenamefont {Howe} \emph {et~al.}}]{Xu2016}%
  \BibitemOpen
  \bibfield  {author} {\bibinfo {author} {\bibfnamefont {S.}~\bibnamefont
  {Xu}}, \bibinfo {author} {\bibfnamefont {F.}~\bibnamefont {Wang}}, \bibinfo
  {author} {\bibfnamefont {C.}~\bibnamefont {Zhu}}, \bibinfo {author}
  {\bibfnamefont {Y.}~\bibnamefont {Meng}}, \bibinfo {author} {\bibfnamefont
  {Y.}~\bibnamefont {Liu}}, \bibinfo {author} {\bibfnamefont {W.}~\bibnamefont
  {Liu}}, \bibinfo {author} {\bibfnamefont {J.}~\bibnamefont {Tang}}, \bibinfo
  {author} {\bibfnamefont {K.}~\bibnamefont {Liu}}, \bibinfo {author}
  {\bibfnamefont {G.}~\bibnamefont {Hu}}, \bibinfo {author} {\bibfnamefont
  {R.}~\bibnamefont {Howe}}, \emph {et~al.},\ }\bibfield  {title} {\bibinfo
  {title} {{Ultrafast Nonlinear Photoresponse of Single-Wall Carbon Nanotubes:
  A Broadband Degenerate Investigation}},\ }\href
  {https://doi.org/10.1039/C6NR00652C} {\bibfield  {journal} {\bibinfo
  {journal} {Nanoscale}\ }\textbf {\bibinfo {volume} {8}},\ \bibinfo {pages}
  {9304} (\bibinfo {year} {2016})}\BibitemShut {NoStop}%
\bibitem [{\citenamefont {Kazaoui}\ \emph {et~al.}(2001)\citenamefont
  {Kazaoui}, \citenamefont {Minami}, \citenamefont {Matsuda}, \citenamefont
  {Kataura},\ and\ \citenamefont {Achiba}}]{Kazaoui2001}%
  \BibitemOpen
  \bibfield  {author} {\bibinfo {author} {\bibfnamefont {S.}~\bibnamefont
  {Kazaoui}}, \bibinfo {author} {\bibfnamefont {N.}~\bibnamefont {Minami}},
  \bibinfo {author} {\bibfnamefont {N.}~\bibnamefont {Matsuda}}, \bibinfo
  {author} {\bibfnamefont {H.}~\bibnamefont {Kataura}},\ and\ \bibinfo {author}
  {\bibfnamefont {Y.}~\bibnamefont {Achiba}},\ }\bibfield  {title} {\bibinfo
  {title} {{Electrochemical tuning of electronic states in single-wall carbon
  nanotubes studied by in situ absorption spectroscopy and ac resistance}},\
  }\href {https://doi.org/10.1063/1.1372208} {\bibfield  {journal} {\bibinfo
  {journal} {Applied Physics Letters}\ }\textbf {\bibinfo {volume} {78}},\
  \bibinfo {pages} {3433} (\bibinfo {year} {2001})}\BibitemShut {NoStop}%
\bibitem [{\citenamefont {Bisri}\ \emph {et~al.}(2017)\citenamefont {Bisri},
  \citenamefont {Shimizu}, \citenamefont {Nakano},\ and\ \citenamefont
  {Iwasa}}]{Bisri2017}%
  \BibitemOpen
  \bibfield  {author} {\bibinfo {author} {\bibfnamefont {S.~Z.}\ \bibnamefont
  {Bisri}}, \bibinfo {author} {\bibfnamefont {S.}~\bibnamefont {Shimizu}},
  \bibinfo {author} {\bibfnamefont {M.}~\bibnamefont {Nakano}},\ and\ \bibinfo
  {author} {\bibfnamefont {Y.}~\bibnamefont {Iwasa}},\ }\bibfield  {title}
  {\bibinfo {title} {{Endeavor of Iontronics: From Fundamentals to Applications
  of Ion-Controlled Electronics}},\ }\href
  {https://doi.org/10.1002/adma.201607054} {\bibfield  {journal} {\bibinfo
  {journal} {Advanced Materials}\ }\textbf {\bibinfo {volume} {29}},\ \bibinfo
  {pages} {1607054} (\bibinfo {year} {2017})}\BibitemShut {NoStop}%
\bibitem [{\citenamefont {Lauret}\ \emph {et~al.}(2003)\citenamefont {Lauret},
  \citenamefont {Voisin}, \citenamefont {Cassabois}, \citenamefont {Delalande},
  \citenamefont {Roussignol}, \citenamefont {Jost},\ and\ \citenamefont
  {Capes}}]{Lauret2003}%
  \BibitemOpen
  \bibfield  {author} {\bibinfo {author} {\bibfnamefont {J.~S.}\ \bibnamefont
  {Lauret}}, \bibinfo {author} {\bibfnamefont {C.}~\bibnamefont {Voisin}},
  \bibinfo {author} {\bibfnamefont {G.}~\bibnamefont {Cassabois}}, \bibinfo
  {author} {\bibfnamefont {C.}~\bibnamefont {Delalande}}, \bibinfo {author}
  {\bibfnamefont {P.}~\bibnamefont {Roussignol}}, \bibinfo {author}
  {\bibfnamefont {O.}~\bibnamefont {Jost}},\ and\ \bibinfo {author}
  {\bibfnamefont {L.}~\bibnamefont {Capes}},\ }\bibfield  {title} {\bibinfo
  {title} {{Ultrafast Carrier Dynamics in Single-Wall Carbon Nanotubes}},\
  }\href {https://doi.org/10.1103/PhysRevLett.90.057404} {\bibfield  {journal}
  {\bibinfo  {journal} {Physical Review Letters}\ }\textbf {\bibinfo {volume}
  {90}},\ \bibinfo {pages} {4} (\bibinfo {year} {2003})}\BibitemShut {NoStop}%
\bibitem [{\citenamefont {Ostojic}\ \emph {et~al.}(2004)\citenamefont
  {Ostojic}, \citenamefont {Zaric}, \citenamefont {Kono}, \citenamefont
  {Strano}, \citenamefont {Moore}, \citenamefont {Hauge},\ and\ \citenamefont
  {Smalley}}]{Ostojic2004}%
  \BibitemOpen
  \bibfield  {author} {\bibinfo {author} {\bibfnamefont {G.~N.}\ \bibnamefont
  {Ostojic}}, \bibinfo {author} {\bibfnamefont {S.}~\bibnamefont {Zaric}},
  \bibinfo {author} {\bibfnamefont {J.}~\bibnamefont {Kono}}, \bibinfo {author}
  {\bibfnamefont {M.~S.}\ \bibnamefont {Strano}}, \bibinfo {author}
  {\bibfnamefont {V.~C.}\ \bibnamefont {Moore}}, \bibinfo {author}
  {\bibfnamefont {R.~H.}\ \bibnamefont {Hauge}},\ and\ \bibinfo {author}
  {\bibfnamefont {R.~E.}\ \bibnamefont {Smalley}},\ }\bibfield  {title}
  {\bibinfo {title} {{Interband Recombination Dynamics in Resonantly Excited
  Single-Walled Carbon Nanotubes}},\ }\href
  {https://doi.org/10.1103/PhysRevLett.92.117402} {\bibfield  {journal}
  {\bibinfo  {journal} {Physical Review Letters}\ }\textbf {\bibinfo {volume}
  {92}},\ \bibinfo {pages} {117402} (\bibinfo {year} {2004})}\BibitemShut
  {NoStop}%
\bibitem [{\citenamefont {Landau}\ \emph {et~al.}(2013)\citenamefont {Landau},
  \citenamefont {Bell}, \citenamefont {Kearsley},\ and\ \citenamefont
  {Pitaevskii}}]{Landau2013}%
  \BibitemOpen
  \bibfield  {author} {\bibinfo {author} {\bibfnamefont {L.}~\bibnamefont
  {Landau}}, \bibinfo {author} {\bibfnamefont {J.}~\bibnamefont {Bell}},
  \bibinfo {author} {\bibfnamefont {M.}~\bibnamefont {Kearsley}},\ and\
  \bibinfo {author} {\bibfnamefont {L.}~\bibnamefont {Pitaevskii}},\ }\href
  {https://books.google.com/books?hl=en\&lr=\&id=jedbAwAAQBAJ\&oi=fnd\&pg=PP1\&dq=landau+and+lifshitz+8\&ots=boLKmO6exr\&sig=223Qrpf69Ju3pozKhv14mD5Ew4U}
  {\emph {\bibinfo {title} {{Electrodynamics of continuous media}}}}\ (\bibinfo
  {year} {2013})\BibitemShut {NoStop}%
\bibitem [{\citenamefont {Shastri}\ \emph {et~al.}(2021)\citenamefont
  {Shastri}, \citenamefont {Tait}, \citenamefont {{Ferreira de Lima}},
  \citenamefont {Pernice}, \citenamefont {Bhaskaran}, \citenamefont {Wright},\
  and\ \citenamefont {Prucnal}}]{Shastri2021}%
  \BibitemOpen
  \bibfield  {author} {\bibinfo {author} {\bibfnamefont {B.~J.}\ \bibnamefont
  {Shastri}}, \bibinfo {author} {\bibfnamefont {A.~N.}\ \bibnamefont {Tait}},
  \bibinfo {author} {\bibfnamefont {T.}~\bibnamefont {{Ferreira de Lima}}},
  \bibinfo {author} {\bibfnamefont {W.~H.}\ \bibnamefont {Pernice}}, \bibinfo
  {author} {\bibfnamefont {H.}~\bibnamefont {Bhaskaran}}, \bibinfo {author}
  {\bibfnamefont {C.~D.}\ \bibnamefont {Wright}},\ and\ \bibinfo {author}
  {\bibfnamefont {P.~R.}\ \bibnamefont {Prucnal}},\ }\bibfield  {title}
  {\bibinfo {title} {{Photonics for artificial intelligence and neuromorphic
  computing}},\ }\href {https://doi.org/10.1038/s41566-020-00754-y} {\bibfield
  {journal} {\bibinfo  {journal} {Nature Photonics}\ }\textbf {\bibinfo
  {volume} {15}},\ \bibinfo {pages} {102} (\bibinfo {year} {2021})}\BibitemShut
  {NoStop}%
\bibitem [{\citenamefont {Krasnikov}\ \emph {et~al.}(2023)\citenamefont
  {Krasnikov}, \citenamefont {Khabushev}, \citenamefont {Gaev}, \citenamefont
  {Bogdanova}, \citenamefont {Iakovlev}, \citenamefont {Lantsberg},
  \citenamefont {Kallio},\ and\ \citenamefont {Nasibulin}}]{Krasnikov2023}%
  \BibitemOpen
  \bibfield  {author} {\bibinfo {author} {\bibfnamefont {D.~V.}\ \bibnamefont
  {Krasnikov}}, \bibinfo {author} {\bibfnamefont {E.~M.}\ \bibnamefont
  {Khabushev}}, \bibinfo {author} {\bibfnamefont {A.}~\bibnamefont {Gaev}},
  \bibinfo {author} {\bibfnamefont {A.~R.}\ \bibnamefont {Bogdanova}}, \bibinfo
  {author} {\bibfnamefont {V.~Y.}\ \bibnamefont {Iakovlev}}, \bibinfo {author}
  {\bibfnamefont {A.}~\bibnamefont {Lantsberg}}, \bibinfo {author}
  {\bibfnamefont {T.}~\bibnamefont {Kallio}},\ and\ \bibinfo {author}
  {\bibfnamefont {A.~G.}\ \bibnamefont {Nasibulin}},\ }\bibfield  {title}
  {\bibinfo {title} {{Machine learning methods for aerosol synthesis of
  single-walled carbon nanotubes}},\ }\href
  {https://doi.org/10.1016/J.CARBON.2022.10.044} {\bibfield  {journal}
  {\bibinfo  {journal} {Carbon}\ }\textbf {\bibinfo {volume} {202}},\ \bibinfo
  {pages} {76} (\bibinfo {year} {2023})}\BibitemShut {NoStop}%
\bibitem [{\citenamefont {Iakovlev}\ \emph {et~al.}(2020)\citenamefont
  {Iakovlev}, \citenamefont {Krasnikov}, \citenamefont {Khabushev},
  \citenamefont {Alekseeva}, \citenamefont {Grebenko}, \citenamefont
  {Tsapenko}, \citenamefont {Zabelich}, \citenamefont {Kolodiazhnaia},\ and\
  \citenamefont {Nasibulin}}]{Iakovlev2020}%
  \BibitemOpen
  \bibfield  {author} {\bibinfo {author} {\bibfnamefont {V.~Y.}\ \bibnamefont
  {Iakovlev}}, \bibinfo {author} {\bibfnamefont {D.~V.}\ \bibnamefont
  {Krasnikov}}, \bibinfo {author} {\bibfnamefont {E.~M.}\ \bibnamefont
  {Khabushev}}, \bibinfo {author} {\bibfnamefont {A.~A.}\ \bibnamefont
  {Alekseeva}}, \bibinfo {author} {\bibfnamefont {A.~K.}\ \bibnamefont
  {Grebenko}}, \bibinfo {author} {\bibfnamefont {A.~P.}\ \bibnamefont
  {Tsapenko}}, \bibinfo {author} {\bibfnamefont {B.~Y.}\ \bibnamefont
  {Zabelich}}, \bibinfo {author} {\bibfnamefont {J.~V.}\ \bibnamefont
  {Kolodiazhnaia}},\ and\ \bibinfo {author} {\bibfnamefont {A.~G.}\
  \bibnamefont {Nasibulin}},\ }\bibfield  {title} {\bibinfo {title}
  {{Fine-tuning of spark-discharge aerosol CVD reactor for single-walled carbon
  nanotube growth: The role of ex situ nucleation}},\ }\href
  {https://doi.org/10.1016/J.CEJ.2019.123073} {\bibfield  {journal} {\bibinfo
  {journal} {Chemical Engineering Journal}\ }\textbf {\bibinfo {volume}
  {383}},\ \bibinfo {pages} {123073} (\bibinfo {year} {2020})}\BibitemShut
  {NoStop}%
\bibitem [{\citenamefont {Khabushev}\ \emph {et~al.}(2020)\citenamefont
  {Khabushev}, \citenamefont {Krasnikov}, \citenamefont {Kolodiazhnaia},
  \citenamefont {Bubis},\ and\ \citenamefont {Nasibulin}}]{Khabushev2020}%
  \BibitemOpen
  \bibfield  {author} {\bibinfo {author} {\bibfnamefont {E.~M.}\ \bibnamefont
  {Khabushev}}, \bibinfo {author} {\bibfnamefont {D.~V.}\ \bibnamefont
  {Krasnikov}}, \bibinfo {author} {\bibfnamefont {J.~V.}\ \bibnamefont
  {Kolodiazhnaia}}, \bibinfo {author} {\bibfnamefont {A.~V.}\ \bibnamefont
  {Bubis}},\ and\ \bibinfo {author} {\bibfnamefont {A.~G.}\ \bibnamefont
  {Nasibulin}},\ }\bibfield  {title} {\bibinfo {title} {{Structure-dependent
  performance of single-walled carbon nanotube films in transparent and
  conductive applications}},\ }\href
  {https://doi.org/10.1016/J.CARBON.2020.01.068} {\bibfield  {journal}
  {\bibinfo  {journal} {Carbon}\ }\textbf {\bibinfo {volume} {161}},\ \bibinfo
  {pages} {712} (\bibinfo {year} {2020})}\BibitemShut {NoStop}%
\bibitem [{\citenamefont {Vorotyntsev}\ \emph {et~al.}(2009)\citenamefont
  {Vorotyntsev}, \citenamefont {Zinovyeva}, \citenamefont {Konev},
  \citenamefont {Picquet}, \citenamefont {Gaillon},\ and\ \citenamefont
  {Rizzi}}]{Vorotyntsev2009}%
  \BibitemOpen
  \bibfield  {author} {\bibinfo {author} {\bibfnamefont {M.~A.}\ \bibnamefont
  {Vorotyntsev}}, \bibinfo {author} {\bibfnamefont {V.~A.}\ \bibnamefont
  {Zinovyeva}}, \bibinfo {author} {\bibfnamefont {D.~V.}\ \bibnamefont
  {Konev}}, \bibinfo {author} {\bibfnamefont {M.}~\bibnamefont {Picquet}},
  \bibinfo {author} {\bibfnamefont {L.}~\bibnamefont {Gaillon}},\ and\ \bibinfo
  {author} {\bibfnamefont {C.}~\bibnamefont {Rizzi}},\ }\bibfield  {title}
  {\bibinfo {title} {{Electrochemical and Spectral Properties of Ferrocene (Fc)
  in Ionic Liquid: 1-Butyl-3-methylimidazolium Triflimide, [BMIM][NTf2].
  Concentration Effects}},\ }\href {https://doi.org/10.1021/JP809095Q}
  {\bibfield  {journal} {\bibinfo  {journal} {Journal of Physical Chemistry B}\
  }\textbf {\bibinfo {volume} {113}},\ \bibinfo {pages} {1085} (\bibinfo {year}
  {2009})}\BibitemShut {NoStop}%
\bibitem [{\citenamefont {Montalb{\'{a}}n}\ \emph {et~al.}(2015)\citenamefont
  {Montalb{\'{a}}n}, \citenamefont {Bol{\'{i}}var}, \citenamefont {{D{\'{i}}az
  Ba{\~{n}}os}},\ and\ \citenamefont {V{\'{i}}llora}}]{Montalban2015}%
  \BibitemOpen
  \bibfield  {author} {\bibinfo {author} {\bibfnamefont {M.~G.}\ \bibnamefont
  {Montalb{\'{a}}n}}, \bibinfo {author} {\bibfnamefont {C.~L.}\ \bibnamefont
  {Bol{\'{i}}var}}, \bibinfo {author} {\bibfnamefont {F.~G.}\ \bibnamefont
  {{D{\'{i}}az Ba{\~{n}}os}}},\ and\ \bibinfo {author} {\bibfnamefont
  {G.}~\bibnamefont {V{\'{i}}llora}},\ }\bibfield  {title} {\bibinfo {title}
  {{Effect of Temperature, Anion, and Alkyl Chain Length on the Density and
  Refractive Index of 1-Alkyl-3-methylimidazolium-Based Ionic Liquids}},\
  }\href {https://doi.org/10.1021/je501091q} {\bibfield  {journal} {\bibinfo
  {journal} {Journal of Chemical and Engineering Data}\ }\textbf {\bibinfo
  {volume} {60}},\ \bibinfo {pages} {1986} (\bibinfo {year}
  {2015})}\BibitemShut {NoStop}%
\bibitem [{\citenamefont {Wu}\ \emph {et~al.}(2018)\citenamefont {Wu},
  \citenamefont {Muntzeck}, \citenamefont {de~los Arcos}, \citenamefont
  {Grundmeier}, \citenamefont {Wilhelm},\ and\ \citenamefont
  {Wagner}}]{Wu2018}%
  \BibitemOpen
  \bibfield  {author} {\bibinfo {author} {\bibfnamefont {X.}~\bibnamefont
  {Wu}}, \bibinfo {author} {\bibfnamefont {M.}~\bibnamefont {Muntzeck}},
  \bibinfo {author} {\bibfnamefont {T.}~\bibnamefont {de~los Arcos}}, \bibinfo
  {author} {\bibfnamefont {G.}~\bibnamefont {Grundmeier}}, \bibinfo {author}
  {\bibfnamefont {R.}~\bibnamefont {Wilhelm}},\ and\ \bibinfo {author}
  {\bibfnamefont {T.}~\bibnamefont {Wagner}},\ }\bibfield  {title} {\bibinfo
  {title} {{Determination of the refractive indices of ionic liquids by
  ellipsometry, and their application as immersion liquids}},\ }\href
  {https://doi.org/10.1364/ao.57.009215} {\bibfield  {journal} {\bibinfo
  {journal} {Applied Optics}\ }\textbf {\bibinfo {volume} {57}},\ \bibinfo
  {pages} {9215} (\bibinfo {year} {2018})}\BibitemShut {NoStop}%
\bibitem [{\citenamefont {Arosa}\ \emph {et~al.}(2018)\citenamefont {Arosa},
  \citenamefont {Algnamat}, \citenamefont {Rodr{\'{i}}guez}, \citenamefont
  {Lago}, \citenamefont {Varela},\ and\ \citenamefont {{De La
  Fuente}}}]{Arosa2018}%
  \BibitemOpen
  \bibfield  {author} {\bibinfo {author} {\bibfnamefont {Y.}~\bibnamefont
  {Arosa}}, \bibinfo {author} {\bibfnamefont {B.~S.}\ \bibnamefont {Algnamat}},
  \bibinfo {author} {\bibfnamefont {C.~D.}\ \bibnamefont {Rodr{\'{i}}guez}},
  \bibinfo {author} {\bibfnamefont {E.~L.}\ \bibnamefont {Lago}}, \bibinfo
  {author} {\bibfnamefont {L.~M.}\ \bibnamefont {Varela}},\ and\ \bibinfo
  {author} {\bibfnamefont {R.}~\bibnamefont {{De La Fuente}}},\ }\bibfield
  {title} {\bibinfo {title} {{Modeling the Temperature-Dependent Material
  Dispersion of Imidazolium-Based Ionic Liquids in the VIS-NIR}},\ }\href
  {https://doi.org/10.1021/acs.jpcc.8b08971} {\bibfield  {journal} {\bibinfo
  {journal} {Journal of Physical Chemistry C}\ }\textbf {\bibinfo {volume}
  {122}},\ \bibinfo {pages} {29470} (\bibinfo {year} {2018})}\BibitemShut
  {NoStop}%
\bibitem [{\citenamefont {Hale}\ and\ \citenamefont {Querry}(1973)}]{Hale1973}%
  \BibitemOpen
  \bibfield  {author} {\bibinfo {author} {\bibfnamefont {G.~M.}\ \bibnamefont
  {Hale}}\ and\ \bibinfo {author} {\bibfnamefont {M.~R.}\ \bibnamefont
  {Querry}},\ }\bibfield  {title} {\bibinfo {title} {{Optical Constants of
  Water in the 200-nm to 200-$\mu$m Wavelength Region}},\ }\href
  {https://doi.org/10.1364/ao.12.000555} {\bibfield  {journal} {\bibinfo
  {journal} {Applied Optics}\ }\textbf {\bibinfo {volume} {12}},\ \bibinfo
  {pages} {555} (\bibinfo {year} {1973})}\BibitemShut {NoStop}%
\end{thebibliography}%

\section*{Supplementary information}\label{suppinfo}
Contact davaizilg@gmail.com or Y.Gladush@skol.tech for Supplementary Information

\begin{acknowledgments}
This work was supported by the Russian Science Foundation grant No. 20-73-10256 (SWCNT synthesis, modulation of characteristic, saturable absorbers). The authors thank the Council on grants (grant number NH-1330.2022.1.3).
\end{acknowledgments}

\end{document}